\documentclass[
  10pt,
  aps,
  pra,
  twocolumn,
  superscriptaddress,
  floatfix
]{revtex4-2}

\usepackage{amsmath}
\usepackage{amssymb}
\usepackage{amsfonts}
\usepackage{bm}
\usepackage{mathtools}
\usepackage{mathrsfs}

\usepackage{physics}

\NewDocumentCommand{\E}{o m}{%
  \IfValueTF{#1}{\mathbb{E}_{#1}}{\mathbb{E}}\left[#2\right]%
}
\NewDocumentCommand{\Ec}{m m}{%
  \mathbb{E}\left(#1 \,\middle|\, #2\right)%
}
\NewDocumentCommand{\V}{o m}{%
  \IfValueTF{#1}{\operatorname{Var}_{#1}}{\operatorname{Var}}\left(#2\right)%
}
\NewDocumentCommand{\Vc}{m m}{%
  \operatorname{Var}\left(#1 \,\middle|\, #2\right)%
}
\NewDocumentCommand{\Cov}{o m m}{%
  \IfValueTF{#1}{\operatorname{Cov}_{#1}}{\operatorname{Cov}}\left(#2, #3\right)%
}
\NewDocumentCommand{\Covc}{o m m m}{%
  \IfValueTF{#1}{\operatorname{Cov}_{#1}}{\operatorname{Cov}}\left(#2, #3 \,\middle|\, #4\right)%
}
\NewDocumentCommand{\Eshot}{m m}{%
  \mathbb{E}_{\mathrm{shots}}\left(#1 \,\middle|\, #2\right)%
}
\NewDocumentCommand{\Vshot}{m m}{%
  \operatorname{Var}_{\mathrm{shots}}\left(#1 \,\middle|\, #2\right)%
}
\NewDocumentCommand{\Covshot}{m m m}{%
  \operatorname{Cov}_{\mathrm{shots}}\left(#1, #2 \,\middle|\, #3\right)%
}

\usepackage{algorithm}
\usepackage{algpseudocode}

\makeatletter
\expandafter\edef\csname ftype@algorithm\endcsname{\the\c@float@type}
\makeatother

\usepackage{aliascnt}

\newcounter{thm}[section]
\renewcommand{\thethm}{\thesection.\arabic{thm}}

\newaliascnt{lemma}{thm}
\newaliascnt{proposition}{thm}
\newaliascnt{corollary}{thm}
\newaliascnt{definition}{thm}

\newenvironment{theorem}[1][]{%
  \par\medskip\refstepcounter{thm}%
  \noindent\textbf{Theorem \thethm\ifx\relax#1\relax\else\ (#1)\fi.}\itshape\ }%
  {\par\medskip}

\newenvironment{lemma}[1][]{%
  \par\medskip\refstepcounter{lemma}%
  \noindent\textbf{Lemma \thelemma\ifx\relax#1\relax\else\ (#1)\fi.}\itshape\ }%
  {\par\medskip}

\newenvironment{proposition}[1][]{%
  \par\medskip\refstepcounter{proposition}%
  \noindent\textbf{Proposition \theproposition\ifx\relax#1\relax\else\ (#1)\fi.}\itshape\ }%
  {\par\medskip}

\newenvironment{corollary}[1][]{%
  \par\medskip\refstepcounter{corollary}%
  \noindent\textbf{Corollary \thecorollary\ifx\relax#1\relax\else\ (#1)\fi.}\itshape\ }%
  {\par\medskip}

  {\par\medskip}

\newenvironment{proof}[1][Proof]{%
  \par\medskip\noindent\textit{#1.}\ }%
  {\hfill$\square$\par\medskip}

\usepackage{hyperref}
\hypersetup{colorlinks=true, linkcolor=blue, citecolor=blue, urlcolor=blue}

\usepackage{cleveref}
\crefname{theorem}{Theorem}{Theorems}
\crefname{lemma}{Lemma}{Lemmas}
\crefname{proposition}{Proposition}{Propositions}
\crefname{corollary}{Corollary}{Corollaries}
\crefname{definition}{Definition}{Definitions}

\begin{document}
% ...

\preprint{APS/123-QED}

\title{Adaptive Protocol for R\'enyi Entropy Estimation}

\author{Mu Niu}
\affiliation{Department of Physics \& Astronomy, University of Waterloo, Ontario, N2L 3G1, Canada}
\affiliation{Institute for Quantum Computing, University of Waterloo, Ontario, N2L 3G1, Canada}
\affiliation{Perimeter Institute for Theoretical Physics, Waterloo, Ontario, N2L 2Y5, Canada}
\email{m5niu@uwaterloo.ca}
\author{Yi Hong Teoh}
\affiliation{Department of Physics \& Astronomy, University of Waterloo, Ontario, N2L 3G1, Canada}
\author{Roger G. Melko}
\affiliation{Department of Physics \& Astronomy, University of Waterloo, Ontario, N2L 3G1, Canada}
\affiliation{Perimeter Institute for Theoretical Physics, Waterloo, Ontario, N2L 2Y5, Canada}
\author{Rajibul Islam}
\affiliation{Department of Physics \& Astronomy, University of Waterloo, Ontario, N2L 3G1, Canada}
\affiliation{Institute for Quantum Computing, University of Waterloo, Ontario, N2L 3G1, Canada}




\date{\today}

\begin{abstract}
The second-order R\'enyi entropy is one of the most important quantities used for characterizing the entanglement structure of states prepared in quantum many-body experiments. 
With access to qubit projective measurement data, it can be estimated via a protocol in which repeated measurements are taken after each of a set of random local unitary operations is applied to the quantum state.
To obtain accurate values for the second R\'enyi entropy on shot-limited devices, care must be taken when choosing the number of different unitary settings to apply, and the respective number of projective measurements to take.
In this paper, we propose an adaptive protocol for estimating the R\'enyi entropy that iteratively optimizes the ratio of basis settings to measurements, using a gradient-based approach and starting from an initial small set of measurements.
Using synthetic data, we demonstrate that for several prototypical states, our method converges to values for this ratio that are comparable to the optimal values obtained by a brute-force grid search, and it exhibits slower growth of the measurement budget required for a fixed target accuracy than fixed-ratio protocols.
Our adaptive protocol therefore promotes efficient estimation of the R\'enyi entropy in finite-size, shot-limited quantum experiments by making effective use of the available measurement budget.
\end{abstract}
\maketitle

\section{Introduction}

The entanglement entropy has emerged as a fundamental quantity for characterizing quantum many-body systems. In states prepared by quantum simulators and similar devices, quantities like the von Neumann entropy and its R\'enyi counterparts are capable of capturing correlations between interacting qubits in a basis-independent way, without reliance on the knowledge of a broken symmetry or local order parameter. This allows the entanglement to provide signatures of a wide variety of exotic phenomena in closed pure systems, for example topological order \cite{LW,KP} or quantum criticality \cite{CC,FM}, where universal quantities can emerge in the scaling of the entanglement entropy. 
More generally, in open, noisy, or otherwise mixed-state systems, recent work has emphasized how some phenomena may be invisible to observables expressed as linear functions of the averaged density matrix, requiring nonlinear information quantities \cite{Fan,Negari,Sang}.
Thus the entanglement entropies and related quantities are increasingly playing a central role in identifying exotic phenomena at the frontier of the field, such as measurement-induced phase transitions \cite{MIPT1,MIPT2}, quantum-error-correcting phases \cite{Choi,Gullans}, and separable and long-range entangled mixed states \cite{Grusdt,Chen,Tarun}.

On an experimental quantum hardware platform that prepares a many-body quantum state, the ability to estimate entanglement entropy and related quantities thus plays a key role. 
Unlike in the case of {\it in silico} simulations, information about an experimental quantum state can only be accessed by measurement.
The resource cost that each measurement requires means that quantities that rely on a brute-force tomographic reconstruction of a density matrix, such as the von Neumann entanglement entropy, will be experimentally inaccessible in a many-qubit system.
However, other entanglement measures, in particular the integer-order R\'enyi entropies $S^{(n)}$, admit direct estimators that are accessible without quantum state tomography.

For an integer $n$, estimating the $n^{\text{th}}$-order R\'enyi entropy $S^{(n)} = \frac{1}{1-n} \log \Tr(\rho^n)$ amounts to evaluating $\Tr(\rho^n)$, which can be expressed as the expectation value of a cyclic permutation operator acting on $n$ identical replicas of the density matrix $\rho$~\cite{CC}.
The simplest case is $n=2$, for which 
existing experimental approaches broadly fall into two classes. The first category relies on preparing two physical copies of the quantum state, and directly exploits the equivalence between the purity $\Tr(\rho^2)$ and the SWAP expectation value, $\Tr(\rho^2) = \Tr[\operatorname{SWAP}(\rho\otimes\rho)]$, where $\operatorname{SWAP} = \sum_{i,j} \ket{ij}\bra{ji}$ in the standard tensor-product basis. Using interference between the two copies, the purity $\Tr(\rho^2)$ can be converted into an observable directly measurable in the experiment~\cite{abanin2012quantum,daley2012entanglement,islam2015entanglement,zhou2024hybrid}. The second class of methods uses only a single copy of the system and relies on the idea that the SWAP-operator expectation value can be reconstructed from the statistics of a single copy of the quantum state evolved under Haar-random unitaries~\cite{elben2019statistical,du2026optimal,brydges2019probing}.

Due to the limited number of qubits available on current atomic, molecular, and optical (AMO) platforms, single-copy protocols offer a significant practical advantage since they avoid the need to prepare multiple identical copies of the system simultaneously.
While there is evidence that multi-copy coherent methods can reduce the sample complexity~\cite{Chen2022}, the SWAP operator requires coherent two-qubit gates acting between copies.  These are considerably more challenging to implement experimentally than single-qubit rotations~\cite{Ransford2026,Bruzewicz2019,Saffman2016,Evered2023}, which are all that is required for the single-copy randomized protocol.
Specifically, the single-copy randomized protocol measures the state under the action of $N_U$ random unitary operations, collecting $N_M$ samples for each unitary. Since the unitaries can be chosen as tensor products of single-qubit random unitaries, this procedure is equivalent to measuring the state in a set of random local bases. From these measurement results, an estimate of $S^{(2)}$ can be constructed as we describe below.

A key subtlety of this method is that the optimal ratio $N_U/N_M$ is state-dependent. For example, a maximally mixed state has exactly basis-independent ideal outcome probabilities, so only the finite-shot empirical frequencies vary across different unitaries. In this case, it is more efficient to sample fewer unitaries but perform more measurements per unitary, making $N_M$ more important than $N_U$. In contrast, a product state can exhibit strong basis dependence: there exist bases in which the state is aligned with a computational basis vector, as well as bases in which the measurement outcomes appear completely random. Accurately capturing this variation across bases requires sampling many different unitaries, making $N_U$ the more important resource in this case. Using a suboptimal parameter ratio could require significantly more experimental resources to reach the same estimation error.

In this work, we build on previous statistical analyses of the randomized measurement protocol to develop an adaptive sampling scheme for $S^{(2)}$. Starting from small initial values of $N_U$ and $N_M$, we iteratively update the ratio $N_U/N_M$ as more measurement data are collected. This update is guided by the information obtained from earlier measurements. All measurement data used in this work are simulated by sampling from computed probabilities. We show that, for representative states covering the maximal, minimal, and intermediate purity values, this adaptive procedure makes the final resource allocation converge to a near-optimal ratio between $N_U$ and $N_M$; we therefore expect the same behavior for states with arbitrary purity. This effectively extends the randomized protocol into a state-blind method with no additional experimental resource overhead.

\begin{algorithm}
\caption{Randomized Measurement Protocol for Purity Estimation (Non-Adaptive)}
\label{alg:randomized_protocol}
\begin{algorithmic}[1]

\State \textbf{Input:} Quantum state $\rho$, subsystem $A$, number of unitaries $N_U$, measurements per unitary $N_M$
\State \textbf{Output:} Estimator of purity $p(\rho, A) = \mathrm{Tr}(\rho_A^2)$ 

\For{$n = 1$ to $N_U$}
    \State Sample a random unitary $U^{(n)}$ from a chosen ensemble
    
    \For{$m = 1$ to $N_M$}
        \State Prepare the state $\rho$
        \State Apply $U^{(n)}$
        \State Measure subsystem $A$ in the computational basis
        \State Record outcome bitstring $s_A^{(n,m)}$
    \EndFor
    
    \State Count the frequency of each outcome bitstring $s_A$
\EndFor

\State Compute the purity estimate using Eq.~\eqref{eq:p_hat estimator}
\State \Return $\widehat{p}$

\end{algorithmic}
\end{algorithm}

\section{Procedure}

\subsection{Randomized Measurement Protocol for R\'enyi Entropy Revisited}
\label{Non-adaptive Protocol Intro}

\subsubsection{Standard Randomized Protocol}

In this section, we briefly review the randomized measurement protocol for estimating the second-order R\'enyi entropy. Consider a system of $N$ interacting qubits and let $A$ be a subsystem containing $N_A$ qubits. The second-order R\'enyi entropy of $A$ is~\cite{brydges2019probing,elben2018renyi}
\begin{equation}
S^{(2)}(\rho, A) = -\log_2\left(\Tr[\rho_A^2]\right),
\label{eq:S2}
\end{equation}
where $\rho_A$ is the reduced density matrix of subsystem $A$, obtained from the full system density matrix $\rho$. From Eq.~\eqref{eq:S2}, $S^{(2)}$ is fully determined by the purity $p(\rho, A) \equiv \Tr[\rho_A^2]$.

In the randomized measurement protocol, the state is measured under the action of $N_U$ random unitaries, with $N_M$ measurement shots collected for each unitary (Algorithm~\ref{alg:randomized_protocol}). Throughout this work, the unitaries are drawn from a local unitary 2-design, i.e., tensor products of independently chosen single-qubit random Clifford unitaries~\cite{Dankert2009ExactApproximateUnitary2Designs}. For this ensemble, the purity of a subsystem $A$ of $N_A$ qubits can be written as~\cite{brydges2019probing}
\begin{equation}
    p(\rho, A) = 2^{N_A}\, \E[U]{\vb{p}_U^{\top} W \vb{p}_U},
    \label{eq:purity_local}
\end{equation}
where $\E[U]{\,\cdot\,}$ denotes the average over the unitary ensemble, $\vb{p}_U$ is the column vector with entries
\begin{equation}
    (\vb{p}_U)_{s} \equiv P_U(s) = \bra{s} U \rho_A U^\dagger \ket{s},
    \label{eq:pU_def}
\end{equation}
collecting the probabilities of measuring each bitstring $s \in \{0,1\}^{N_A}$ on subsystem $A$ after the state is transformed by $U$, and the Hamming kernel $W$ is defined entrywise by
\begin{equation}
    W_{ss'} \equiv (-2)^{-D[s, s']},
    \label{eq:kernel}
\end{equation}
where $D[s, s']$ is the Hamming distance between $s$ and $s'$, i.e., the number of bits on which the two strings differ.

\subsubsection{Hamming Kernel Formulation}

To write the estimator in a more compact operator form, we define the diagonal operator acting on two copies of subsystem $A$,
\begin{equation}
    X \equiv 2^{N_A} \sum_{s, s'} W_{ss'}\, \ket{s, s'}\bra{s, s'},
    \label{eq:X}
\end{equation}
where $\ket{s, s'} \equiv \ket{s} \otimes \ket{s'}$, and its unitary transformed version
\begin{equation}
    X_U \equiv U^{\dagger \otimes 2}\, X\, U^{\otimes 2}, \qquad U^{\otimes 2} \equiv U \otimes U.
    \label{eq:XU}
\end{equation}
We also write 
\begin{equation}
    \mathcal{X}_U \equiv \Tr[X_U \rho_A^{\otimes 2}]
\end{equation}
to emphasize that $\mathcal{X}_U$ is the real scalar obtained from the Hilbert--Schmidt inner product between $X_U$ and $\rho_A^{\otimes 2}$.

The key identity behind Eq.~\eqref{eq:purity_local} is that the ensemble average of $X_U$ over a unitary 2-design equals the SWAP operator~\cite{elben2019statistical,du2026optimal},
\begin{equation}
    \E[U]{X_U} = \E[U]{U^{\dagger \otimes 2} X U^{\otimes 2}} = \operatorname{SWAP}.
    \label{eq:XU=SWAP}
\end{equation}
Using this relation, Eq.~\eqref{eq:purity_local} follows from a few lines of algebra.

\subsubsection{Experimental Estimator of Purity}

In experiments, the probabilities $P_U(s)$ are not accessible and they need to be estimated from measurement records. Simply replacing each $P_U(s)$ in Eq.~\eqref{eq:purity_local} by the empirical frequency $ c_U(s)/N_M$, where $c_U(s)$ is the count for string $s$ under unitary $U$, yields a biased estimator since purity is quadratic in outcome probabilities, and in general squaring an unbiased estimator for $P_U(s)$ will be a biased estimator of $P_U(s)^2$. An exactly unbiased estimator for $\mathcal{X}_U$ is given by~\cite{Elben2023Toolbox}
\begin{equation}
    \label{eq: Purity Estimator Kernel Form}
    \widehat{\mathcal{X}_U} \equiv \frac{2^{N_A}}{(N_M)^{\underline{2}}} \sum_{i,j \,|\, i \neq j} W_{s_i s_j},
\end{equation}
where $i$ and $j$ are sample indices labeling the $i^{\text{th}}$ and $j^{\text{th}}$ measurement shots, while $s_i$ and $s_j$ are the corresponding outcome strings. The sum $\sum_{i,j | i \neq j}$ denotes that we are summing over both indices $i$ and $j$, subject to the constraint $i \neq j$. The notation $n^{\underline{k}} = n!/(n-k)!$ denotes the falling factorial of $n$.

The quantity $\widehat{\mathcal{X}_U}$ is an unbiased estimator of $\mathcal{X}_U$ in the sense that $ \Eshot{\widehat{\mathcal{X}_U}}{U}  = \mathcal{X}_U$, where $\Eshot{\,\cdot\,}{U}$ denotes the average over the sample space of all $N_M$-shot outcome sequences, $\Omega = \{(s_1, \dots, s_{N_M}) : s_i \in \{0,1\}^{N_A}\}$ for a given unitary $U$. It can also be viewed as an unbiased estimator for purity in the sense that
\begin{equation}
\label{eq:unbiasedness}
    \E[U]{\Eshot{\widehat{\mathcal{X}_U}}{U}} = 
    \E[U]{\mathcal{X}_U} = p(\rho,A).
\end{equation}
In outcome space, we may write the estimator in terms of counts only,
\begin{equation}
    \widehat{\mathcal{X}_U} = \frac{2^{N_A}}{(N_M)^{\underline{2}}} \left(\vb{c}_U^{\top} W \vb{c}_U - N_M\right),
    \label{eq:estimator for TrXU count space}
\end{equation}
where the count vector $\vb{c}_U$, defined by $(\vb{c}_U)_{s_A} \equiv c_U(s_A)$, collects the counts for each possible outcome string $s_A$. We provide a self-contained proof of Eq.~\eqref{eq:unbiasedness} and Eq.~\eqref{eq:estimator for TrXU count space} in Appendix~\ref{Unbiased Estimator for Purity} based on results in Refs.~\cite{vermersch2018unitary,Elben2023Toolbox}.

Since the average of unbiased estimators is still unbiased, in the actual implementation of the protocol we use
\begin{equation}
    \widehat{p} \equiv \frac{1}{N_U} \sum_{r=1}^{N_U} \widehat{\mathcal{X}_{U_r}}
    \label{eq:p_hat estimator}
\end{equation}
as the estimator for purity. It is unbiased in the sense that $\E[U_1, \dots, U_{N_U}]{\Eshot{\widehat{p}}{U_1 \dots U_{N_U}}} = p(\rho, A)$. We remark that the unbiasedness is exact when the parameters $N_U$ and $N_M$ are fixed across all trials; in the adaptive scheme, where these parameters become data-dependent, we do not claim that the final estimator remains exactly unbiased. However, since the final allocation $(N_U, N_M)$ concentrates across independent trials for any given state (Fig.~\ref{fig:Demonstration of Protocol}), the estimator effectively behaves as a fixed-parameter estimator at the concentrated allocation, and we expect any deviation from unbiasedness to be negligible. We also do not claim that taking the logarithm of $\widehat{p}$ yields an exactly unbiased estimator of $S^{(2)}$; in practice this bias is negligible, and the same plug-in estimator has been used in previous work to estimate $S^{(2)}$~\cite{brydges2019probing}.

\subsection{Adaptive Scheme}

\subsubsection{Total Variance and Variance Decomposition}

In the standard randomized measurement protocol outlined in Sec.~\ref{Non-adaptive Protocol Intro}, the numbers of unitaries $N_U$ and measurements per unitary $N_M$ are fixed input parameters specified \emph{a priori}. In the adaptive scheme, by contrast, we begin with an initial set of unitaries and measurements, denoted by $N_U^{(0)}$ and $N_M^{(0)}$, and iteratively update $N_U$ and $N_M$ throughout the protocol. The adaptive process terminates when the available experimental resource budget is exhausted.

At each iteration, the decision to increase either $N_U$ or $N_M$ is determined by a gradient-based estimate of the statistical error. The total variance of the purity estimate comes from two sources: the finite number of sampled unitaries and the finite number of measurements performed given a fixed unitary. Using the law of total variance~\cite{blitzstein_hwang_2019_probability}, it can be decomposed as
\begin{equation}
\label{eq:total variance}
\V{\widehat{p}} = \frac{V_U + V_{\mathrm{shot}}(N_M)}{N_U},
\end{equation}
where 
\begin{equation}
    \label{eq: V_U}
    V_U \equiv \V[U]{\mathcal{X}_U},
\end{equation}
which represents the variance of $\mathcal{X}_U$ across the sampled unitaries and depends only on the state $\rho$, and
\begin{equation}
    V_{\mathrm{shot}}(N_M) \equiv \E[U]{\Vshot{\widehat{\mathcal{X}_U}}{U}}.
\end{equation}
Here, we first compute the variance of $\widehat{\mathcal{X}_U}$ over shots conditional on $U$, and then average over the unitary ensemble.

According to Eq.~\eqref{eq:total variance}, increasing $N_U$ reduces both variance contributions. Intuitively, sampling an additional unitary that is close to some existing unitary is effectively measuring that existing unitary twice. Using U-statistics~\cite{Hoeffding1948}, we can further write the shot noise contribution as
\begin{equation}
\label{eq:shot noise variance exact}
    V_{\mathrm{shot}}(N_M) = \frac{4(N_M-2)\,\bar\zeta_1 + 2\,\bar\zeta_2}{(N_M)^{\underline{2}}},
\end{equation}
where 
\begin{equation}\label{eq:zeta coefficients}
\begin{aligned}
\zeta_1(U) &= 4^{N_A} \sum_{s_1, s_2, s_3} W_{s_1 s_2}\, W_{s_1 s_3}\, P_U(s_1) P_U(s_2) P_U(s_3) \\
& - \mathcal{X}_U^2,\\
\zeta_2(U) &= 4^{N_A} \sum_{s_1, s_2} W_{s_1 s_2}^2\, P_U(s_1) P_U(s_2) - \mathcal{X}_U^2,
\end{aligned}
\end{equation}
and 
\begin{equation}
    \bar\zeta_1 = \E[U]{\zeta_1(U)}, \quad \bar\zeta_2 = \E[U]{\zeta_2(U)}.
\end{equation}
Note that the coefficients $\zeta_1$ and $\zeta_2$ depend on the unitary $U$ and implicitly on the state $\rho$, while $\bar\zeta_1$ and $\bar\zeta_2$ are their averages over the unitary ensemble and depend only on the state. The coefficients $\zeta_1(U)$ and $\zeta_2(U)$ are nonnegative: the former is the variance of a conditional expectation, and the latter is a variance. A proof of Eq.~\eqref{eq:total variance} and Eq.~\eqref{eq:shot noise variance exact} is given in Appendix~\ref{app:Variance Analysis of Randomized Protocol}.

\subsubsection{Gradients of Total Error}

We define the total error as
\begin{equation}
    \label{eq:total error}
    \epsilon(N_U, N_M) \equiv \sqrt{\V{\widehat{p}}} = \sqrt{\frac{V_U + V_{\mathrm{shot}}(N_M)}{N_U}}.
\end{equation}

By direct computation of derivatives of Eq.~\eqref{eq:total error} with respect to $N_U$ and $N_M$, we obtain
\begin{equation}
    \label{eq: derivative with N_U}
    \frac{1}{N_M} \abs{\pdv{\epsilon(N_U, N_M)}{N_U}} = \frac{\epsilon(N_U, N_M)}{2N_U N_M}
\end{equation}
and
\begin{equation}
    \label{eq: derivative with N_M}
    \frac{1}{N_U} \abs{\pdv{\epsilon(N_U, N_M)}{N_M}} = \frac{1}{2 \epsilon(N_U, N_M) N_U^2} \abs{\pdv{V_{\mathrm{shot}}(N_M)}{N_M}},
\end{equation}
where
\begin{equation}
    \label{eq: derivative of V_shot with N_M}
    \begin{split}
    \abs{\pdv{V_{\mathrm{shot}}(N_M)}{N_M}} = \frac{1}{\left((N_M)^{\underline{2}}\right)^2} \Bigl[\, 4\bar\zeta_1 N_M^2 &- \left(16\bar\zeta_1 - 4\bar\zeta_2\right) N_M \\
    &+ \left(8\bar\zeta_1 - 2\bar\zeta_2\right) \Bigr].
    \end{split}
\end{equation}
For the regime $N_M \geq 4$ used throughout this work, we have $\pdv{V_{\mathrm{shot}}(N_M)}{N_M} \leq 0$. To see this, rewrite the numerator in Eq.~\eqref{eq: derivative of V_shot with N_M} as $4\bar\zeta_1(N_M^2 - 4N_M + 2) + 2\bar\zeta_2(2N_M - 1)$ and use $\bar\zeta_1,\bar\zeta_2 \geq 0$. Hence the absolute value can be safely replaced by an overall minus sign.

Quantities in Eq.~\eqref{eq: derivative with N_U} and Eq.~\eqref{eq: derivative with N_M} are relevant for the adaptive decision because they are proportional to the change in total error $\epsilon$ when $N_U$ or $N_M$ is increased by one unit, with the derivatives understood as a continuous approximation to these integer increments. The factors $\frac{1}{N_M}$ and $\frac{1}{N_U}$ in these equations are not part of the gradients; they normalize by the measurement cost of each direction, since increasing $N_U$ by one unit increases the total resource $N_U N_M$ by $N_M$, while increasing $N_M$ by one unit increases it by $N_U$. In practice, we are only interested in the relative magnitude of the two gradients, so we can define the effective descents by multiplying both Eq.~\eqref{eq: derivative with N_U} and Eq.~\eqref{eq: derivative with N_M} by $2\epsilon(N_U,N_M) N_U^2 N_M$:
\begin{equation}
    \label{eq:Loss_U}
    L_U = \epsilon(N_U, N_M)^2 N_U = V_U + V_{\mathrm{shot}}(N_M)
\end{equation}
and
\begin{equation}
    \label{eq:Loss_M}
    L_M = \alpha N_M \abs{\pdv{V_{\mathrm{shot}}(N_M)}{N_M}},
\end{equation}
where $\alpha$ is an empirical constant; larger values of $\alpha$ bias the adaptive decision toward increasing $N_M$. We set $\alpha = 0.6$ by optimizing the resource scaling at small subsystem sizes ($N_A = 1$ to $4$). With this choice of $\alpha$, the adaptive protocol drives the final allocation $(N_U, N_M)$ into the desired regime and exhibits a lower resource-scaling exponent than every fixed-ratio protocol on all states tested (see Section~\ref{Results}). We do not, however, claim that this choice is optimal.

\subsubsection{Experimental Estimators of Effective Descents}

Just like the purity estimator, the effective descents $L_U$ and $L_M$ are also state-dependent quantities that are not directly accessible in experiments. We can, however, construct unbiased estimators for them using only experimental counts. We note that $V_{\mathrm{shot}}(N_M)$ (Eq.~\eqref{eq:shot noise variance exact}) and $ \abs{\pdv{V_{\mathrm{shot}}(N_M)}{N_M}}$ (Eq.~\eqref{eq: derivative of V_shot with N_M}) are both linear combinations of $\bar\zeta_1$ and $\bar\zeta_2$. Hence unbiased estimators for $V_{\mathrm{shot}}(N_M)$ and $ \abs{\pdv{V_{\mathrm{shot}}(N_M)}{N_M}}$ follow from unbiased estimators for $\bar\zeta_1$ and $\bar\zeta_2$, and together with the unbiased estimator for $V_U$ these yield unbiased estimators for $L_U$ and $L_M$.

The unbiased estimator for $V_U$ is given by
\begin{equation}
    \label{eq:V_U estimator}
    \widehat{V_U}
    = \frac{1}{N_U}\sum_{r=1}^{N_U} \widehat{\mathcal{X}_{U_r}^{\,2}}
    - \frac{1}{(N_U)^{\underline{2}}} \sum_{r \neq r'} \widehat{\mathcal{X}_{U_r}}\;\widehat{\mathcal{X}_{U_{r'}}},
\end{equation}
where $\widehat{\mathcal{X}_{U}^{\,2}}$ is an unbiased estimator for $\mathcal{X}_{U}^{\,2}$, which can be represented in count space as
\begin{equation}
    \label{eq:estimator for TrXU^2 count space}
    \widehat{\mathcal{X}_{U}^{\,2}} = \frac{4^{N_A}}{(N_M)^{\underline{4}}} \left[A_U^2 - 4T_U - 2\left(\vb{c}_U^{\top} W_2 \vb{c}_U - N_M\right)\right],
\end{equation}
where 
\begin{equation}
    \label{eq:A_U}
    A_U \equiv \vb{c}_U^{\top} W \vb{c}_U - N_M,
\end{equation}
and 
\begin{equation}
    \label{eq:T_U}
    T_U \equiv \sum_s c_U(s)\,\left[ \left( \left( W \vb{c}_U \right)_{s} - 1 \right)^2 - \left( W_2 \vb{c}_U \right)_{s} + 1 \right].
\end{equation}
The matrix $W_2$ is the entrywise squared version of the Hamming kernel $(W_2)_{ss'} \equiv W_{ss'}^2$. Since we need $(N_M)^{\underline{4}}$ to compute $\widehat{\mathcal{X}_{U}^{\,2}}$, we must start with $N_M \geq 4$. Similarly, since we need $(N_U)^{\underline{2}}$ to compute $ \widehat{V_U}$, we must start with $N_U \geq 2$.

The unbiased estimators for $\bar\zeta_1$ and $\bar\zeta_2$ are given by
\begin{equation}
    \label{eq:bar_zeta_1 estimator}
    \widehat{\bar\zeta_1} = \frac{1}{N_U} \sum_{r=1}^{N_U} \left(\frac{4^{N_A}}{(N_M)^{\underline{3}}} T_{U_r} - \widehat{\mathcal{X}_{U_r}^{\,2}}\right),
\end{equation}
and
\begin{equation}
    \label{eq:bar_zeta_2 estimator}
    \widehat{\bar\zeta_2} = \frac{1}{N_U} \sum_{r=1}^{N_U} \left(\frac{4^{N_A}}{(N_M)^{\underline{2}}} \left(\vb{c}_{U_r}^{\top} W_2 \vb{c}_{U_r} - N_M\right) - \widehat{\mathcal{X}_{U_r}^{\,2}}\right),
\end{equation}
respectively, in the sense that $\E[U_1, \dots, U_{N_U}]{\Eshot{\widehat{\bar\zeta_i}}{U_1 \dots U_{N_U}}} = \bar\zeta_i$ for $i=1,2$. We provide a proof of Eqs.~\eqref{eq:V_U estimator}, \eqref{eq:bar_zeta_1 estimator} and \eqref{eq:bar_zeta_2 estimator} in Appendix~\ref{app:Estimators of Quantities in Adaptive Decision}.

It follows that we may write the unbiased estimators for $L_U$ and $L_M$ as
\begin{equation}
    \label{eq:Loss estimators}
    \begin{split}
    \widehat{L_U} &= \widehat{V_U} + \frac{4(N_M-2)\,\widehat{\bar\zeta_1} + 2\,\widehat{\bar\zeta_2}}{(N_M)^{\underline{2}}}, \\
    \widehat{L_M} &= \alpha N_M \frac{4\widehat{\bar\zeta_1}\, N_M^2 - \bigl(16\widehat{\bar\zeta_1} - 4\widehat{\bar\zeta_2}\bigr) N_M
        + \bigl(8\widehat{\bar\zeta_1} - 2\widehat{\bar\zeta_2}\bigr)}{\left((N_M)^{\underline{2}}\right)^2},
    \end{split}
\end{equation}
obtained from Eq.~\eqref{eq:shot noise variance exact} and Eq.~\eqref{eq: derivative of V_shot with N_M} with $\bar\zeta_i$ replaced by their unbiased estimators $\widehat{\bar\zeta_i}$. Both expressions are linear in these unbiased ingredients.

At each iteration, we compare $\widehat{L_U}$ and $\widehat{L_M}$ and update the parameter corresponding to the larger descent direction. Specifically, if $\widehat{L_U} \geq \widehat{L_M}$, additional random unitaries are sampled; otherwise, additional measurements are performed for the existing set of unitaries. This procedure is repeated until the experimental budget $N_{\mathrm{tot}}$ is saturated by $N_U N_M$. As the budget approaches saturation, the remaining resource $r = N_{\mathrm{tot}} - N_U N_M$ may no longer suffice for a full step of $\Delta N_U$ unitaries or $\Delta N_M$ additional shots per unitary. In this case, we take the largest partial step that still fits, so that as much of the remaining budget as possible is used: since each new unitary costs $N_M$ measurements, $N_U$ is increased by $\min\!\left( \Delta N_U,\; \lfloor r / N_M \rfloor \right)$, and since one additional shot on every existing unitary costs $N_U$ measurements in total, $N_M$ is increased by $\min\!\left( \Delta N_M,\; \lfloor r / N_U \rfloor \right)$, where $\lfloor \cdot \rfloor$ denotes the floor function. When even a unit increment in the chosen direction no longer fits within the remaining budget, the protocol terminates. At the beginning of the protocol, the adaptive decision can be dominated by shot noise. For example, there is a possibility that for a few initial unitaries that we drew, the state looks almost identical. The protocol would then keep increasing $N_M$ and never sample new unitaries. To prevent this, we introduce a warm-up anti-cascade guard. If $N_U$ is below a guard threshold $N_U^{\mathrm{guard}}$, we force the protocol to increase $N_U$ for at most $c_{\max}$ consecutive iterations that increase $N_M$. The complete adaptive protocol is summarized in Algorithm~\ref{alg:adaptive_protocol}. 

\begin{algorithm}
\caption{Adaptive Randomized Measurement Protocol for Purity Estimation}
\label{alg:adaptive_protocol}
\begin{algorithmic}[1]

\State \textbf{Input:} Quantum state $\rho$, subsystem $A$, initial parameters $N_U^{(0)} \geq 2$, $N_M^{(0)} \geq 4$, step sizes $\Delta N_U, \Delta N_M$, resource budget $N_{\mathrm{tot}}$, empirical constant $\alpha$
\State \textbf{Guard parameters:} $N_U^{\mathrm{guard}}=20$, $c_{\max}=8$
\State \textbf{Output:} Estimate $\widehat{p}$ of the purity $p(\rho, A)$

\State Sample $N_U^{(0)}$ random unitaries $\{U_r\}$
\State For each unitary, perform $N_M^{(0)}$ measurements
\State $N_U \gets N_U^{(0)}$, $N_M \gets N_M^{(0)}$
\State $c \gets 0$

\While{$N_U N_M < N_{\mathrm{tot}}$}
    \State $r \gets N_{\mathrm{tot}} - N_U N_M$ \Comment{remaining budget}
    \State Evaluate $\widehat{\mathcal{X}_{U_r}}$ for each existing unitary $U_r$ using Eq.~\eqref{eq:estimator for TrXU count space}

    \State Estimate $\widehat{V_U}$, $\widehat{\bar\zeta_1}$ and $\widehat{\bar\zeta_2}$ from the counts using Eqs.~\eqref{eq:V_U estimator}, \eqref{eq:bar_zeta_1 estimator} and \eqref{eq:bar_zeta_2 estimator}

    \State Compute the estimators $\widehat{L_U}$ and $\widehat{L_M}$ using Eq.~\eqref{eq:Loss estimators}, with the estimated shot variance and its derivative magnitude clipped at zero

    \State $g \gets (N_U < N_U^{\mathrm{guard}} \ \text{and}\ c \geq c_{\max})$ \Comment{warm-up anti-cascade guard}
    \If{$g \lor (\widehat{L_U} \geq \widehat{L_M})$}
        \State \textbf{(Increase number of unitaries)}
        \State $n \gets \min\!\left( \Delta N_U,\; \lfloor r / N_M \rfloor \right)$
        \If{$n = 0$}
            \State \textbf{break} \Comment{no further unitary fits in the budget}
        \EndIf
        \State Sample $n$ new random unitaries
        \State For each new unitary, perform $N_M$ measurements
        \State $N_U \gets N_U + n$
        \State $c \gets 0$
    \Else
        \State \textbf{(Increase measurements per unitary)}
        \State $m \gets \min\!\left( \Delta N_M,\; \lfloor r / N_U \rfloor \right)$
        \If{$m = 0$}
            \State \textbf{break} \Comment{a uniform extra shot per unitary does not fit the budget}
        \EndIf
        \State For each existing unitary, perform $m$ additional measurements
        \State $N_M \gets N_M + m$
        \State $c \gets c + 1$
    \EndIf
\EndWhile

\State Compute $\widehat{p}$ using Eq.~\eqref{eq:p_hat estimator}
\State \Return $\widehat{p}$ 

\end{algorithmic}
\end{algorithm}

\section{Results}\label{Results}

\subsection{Demonstration of Adaptive Resource Allocation}

\begin{figure}
    \centering
    \includegraphics[width=1\linewidth]{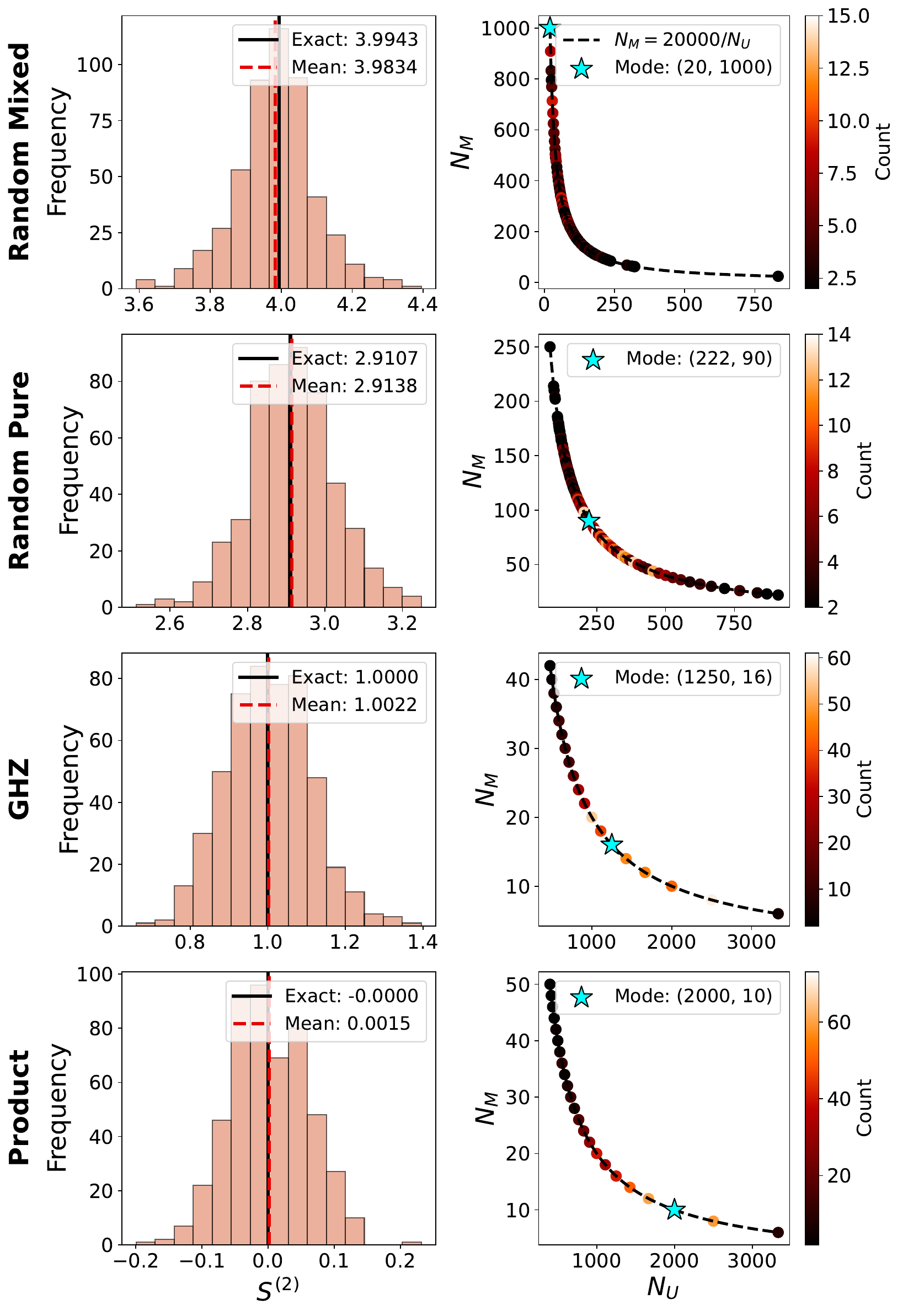}
    \caption{Demonstration of the adaptive protocol for estimating the second R\'enyi entropy $S^{(2)}$ on the half-system partition of four classes of 8-qubit states: random mixed states (whose reduced states are nearly maximally mixed), Haar-random pure states (whose reduced states are nearly maximally mixed at large system sizes), GHZ states, and product states. Results are collected from 500 independent adaptive trials with a total experimental budget of $N_{\mathrm{tot}} = 20000$, starting from $N_U^{(0)} = 2$ and $N_M^{(0)} = 4$ with step sizes $\Delta N_U = \Delta N_M = 2$. \textbf{Left column:} Histograms of the final estimated $S^{(2)}$. The solid vertical line marks the exact entropy value, and the dashed vertical line denotes the mean estimate over the 500 trials. The estimates are approximately normally distributed around the exact value for all four classes of states tested. \textbf{Right column:} Final resource allocation in the $(N_U, N_M)$ plane, where the color scale indicates the number of trials ending at that particular allocation. The cyan star indicates the mode of the allocation among the 500 trials. The protocol successfully adapts the $N_U/N_M$ ratio to the state's properties: it shifts toward larger $N_M$ ($N_U/N_M < 1$) for the nearly maximally mixed reduced states of random mixed states, and toward larger $N_U$ ($N_U/N_M \gg 1$) for GHZ states (whose half-system partition has purity $p = 0.5$) and product states. The dashed curve plots $N_M = 20000/N_U$. All trials run until the measurement budget is saturated.}
    \label{fig:Demonstration of Protocol}
\end{figure}

To demonstrate the adaptive protocol (Algorithm~\ref{alg:adaptive_protocol}), we numerically simulated 500 independent runs of the protocol on the half-partitions of several 8-qubit states spanning a wide range of subsystem purities: random mixed states, Haar-random pure states, Greenberger–Horne–Zeilinger (GHZ) states, and product states. The random mixed states are drawn from the Hilbert--Schmidt ensemble, i.e., $\rho = XX^\dagger/\Tr(XX^\dagger)$ with $X$ a complex Ginibre random matrix, generated using QuTiP's \texttt{rand\_dm} function~\cite{Johansson2012QuTiP}. For Haar-random pure states, the reduced density matrix of a half-partition is nearly maximally mixed for large Hilbert-space dimension as a consequence of the concentration of measure~\cite{plodzien2025scrambling,Hayden2006}. In each simulated run, the measurement outcomes are generated by sampling from the exact outcome probability distributions computed from the corresponding density matrices. Figure~\ref{fig:Demonstration of Protocol} shows the distribution of the final $S^{(2)}$ estimates together with a scatter plot of the final $(N_U, N_M)$ allocations.

The adaptive procedure consistently approaches near-optimal resource allocation. We observe that the final allocated ratio $N_U/N_M$ increases with the purity of the reduced density matrix. In particular, for the product state the final allocation is concentrated around $N_U/N_M \sim 10^2$, while for the nearly maximally mixed reduced states of random mixed states, the final allocation is concentrated around $N_U/N_M \sim 10^{-1}$. 
This is consistent with our grid search plot in Fig.~\ref{fig:Comparison with Grid}, which indicates that for the product state the highest $N_U/N_M$ ratio gives the smallest error bar, and for the random mixed state the lowest $N_U/N_M$ ratio gives the smallest error bar.

\subsection{Demonstration of Estimating Multiple Subsystem Purities}

When the unitary ensemble in Algorithm~\ref{alg:randomized_protocol} is chosen as a tensor product of local random unitaries, we can simultaneously estimate the purity of multiple subsystems from a single set of measurement outcomes. This capability is particularly useful for verifying the scaling of entanglement entropy, e.g., distinguishing between area-law and volume-law states. Instead of only measuring qubits in the subsystem, we measure every qubit of the whole system, obtaining a full bitstring $s$. For any subsystem $A$ of interest, we simply retain the bits that belong to $A$ and discard the rest; the resulting substring is then used to compute the purity of $A$. A single set of measurements therefore gives access to the purity of every subsystem at once. The same property holds for the adaptive protocol of Algorithm~\ref{alg:adaptive_protocol}, with the small subtlety that we must choose a particular target subsystem $A_{\mathrm{target}}$ whose data are used to compute the variance estimates $\widehat{V_U}$, $\widehat{\bar\zeta_1}$ and $\widehat{\bar\zeta_2}$ that determine the adaptive decisions.

\begin{figure}
    \centering
    \includegraphics[width=1\linewidth]{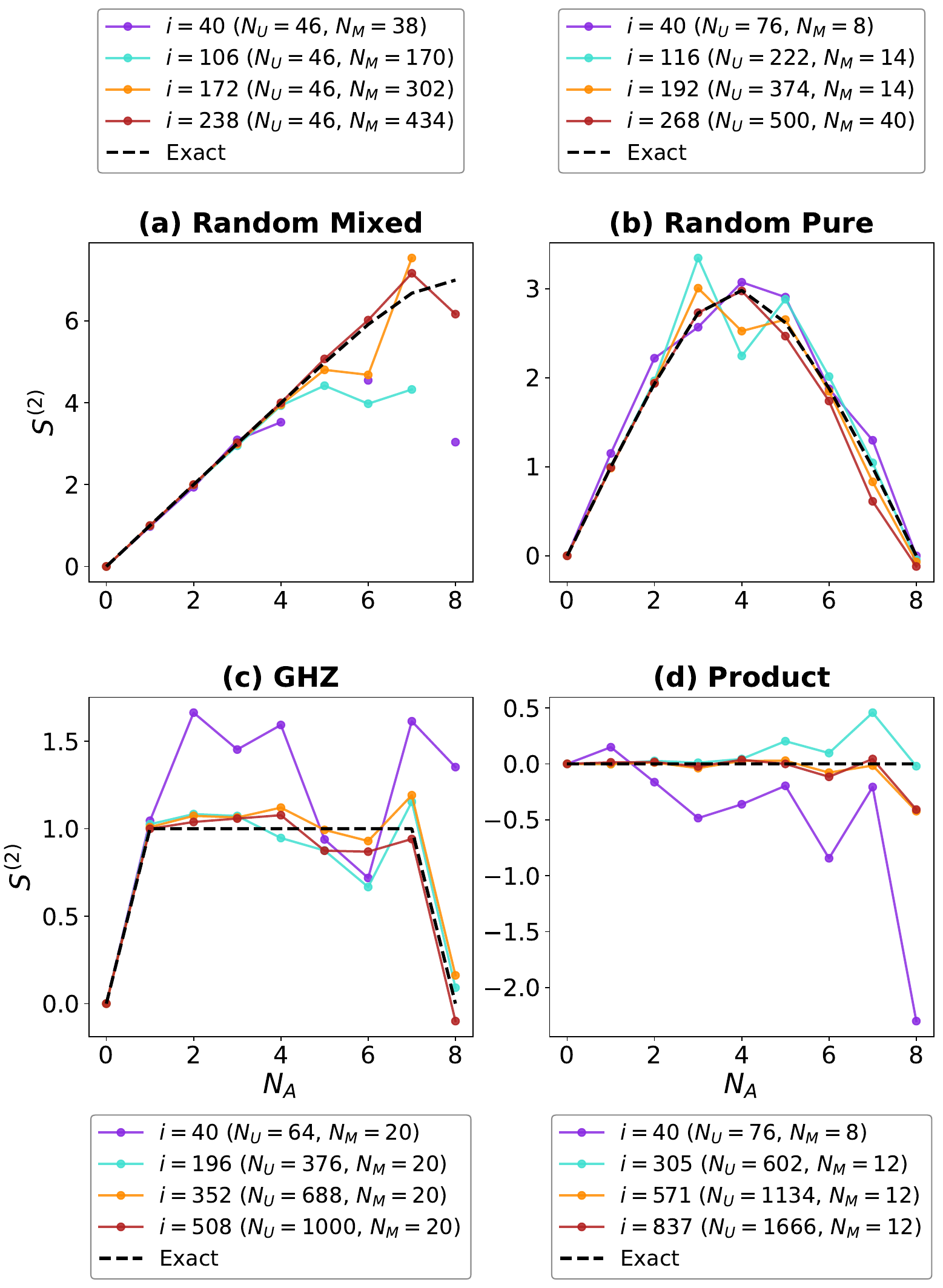}
    \caption{Simultaneous estimation of the second R\'enyi entropy for subsystems $A$ of size $N_A = 1$ to $N$ using the adaptive protocol. The full system consists of $N=8$ qubits, and the target subsystem $A_{\mathrm{target}}$ used to determine the adaptive resource allocation is chosen to be the first four qubits, so that $N_A=4$. The total experimental budget is fixed to $N_{\mathrm{tot}} = 20000$; the protocol runs until the budget is saturated. For each state, the estimated entropies for all subsystem sizes $N_A$ at a given iteration are shown as a single curve. Four representative iterations are plotted for each case. The shown iteration indices $i$ are chosen to be evenly spaced between iteration $i=40$ and the final iteration at which the measurement budget is saturated. Four types of quantum states are considered: (a) random mixed states, (b) Haar-random pure states, (c) GHZ states, and (d) product states. A missing data point indicates that the corresponding $S^{(2)}$ estimate at that subsystem size and iteration is not finite.
}
    \label{fig:all subsystem estimation}
\end{figure}

Figure~\ref{fig:all subsystem estimation} demonstrates the simultaneous estimation of R\'enyi entropies for all contiguous subsystems of the 1D chain starting from the boundary. The subsystem $A$ is taken to be the first $N_A$ qubits, and the $N_A=0$ entropy is set to zero. The total measurement budget is $N_{\mathrm{tot}} = 20000$. As the number of iterations grows, the estimated entropy curve steadily approaches the exact curve. The improvement is especially pronounced for larger $N_A$, because the sample complexity of the protocol scales with the subsystem size~\cite{huang2020predicting}: at fixed total system size $N$, a larger subsystem requires a larger resource $N_U N_M$ to reach the same accuracy, so each increase of the budget yields a more visible improvement for larger subsystems.

\subsection{Comparison with Non-Adaptive Method}
\label{Comparison with Non-Adaptive Method}

\begin{figure}
    \centering
    \includegraphics[width=1\linewidth]{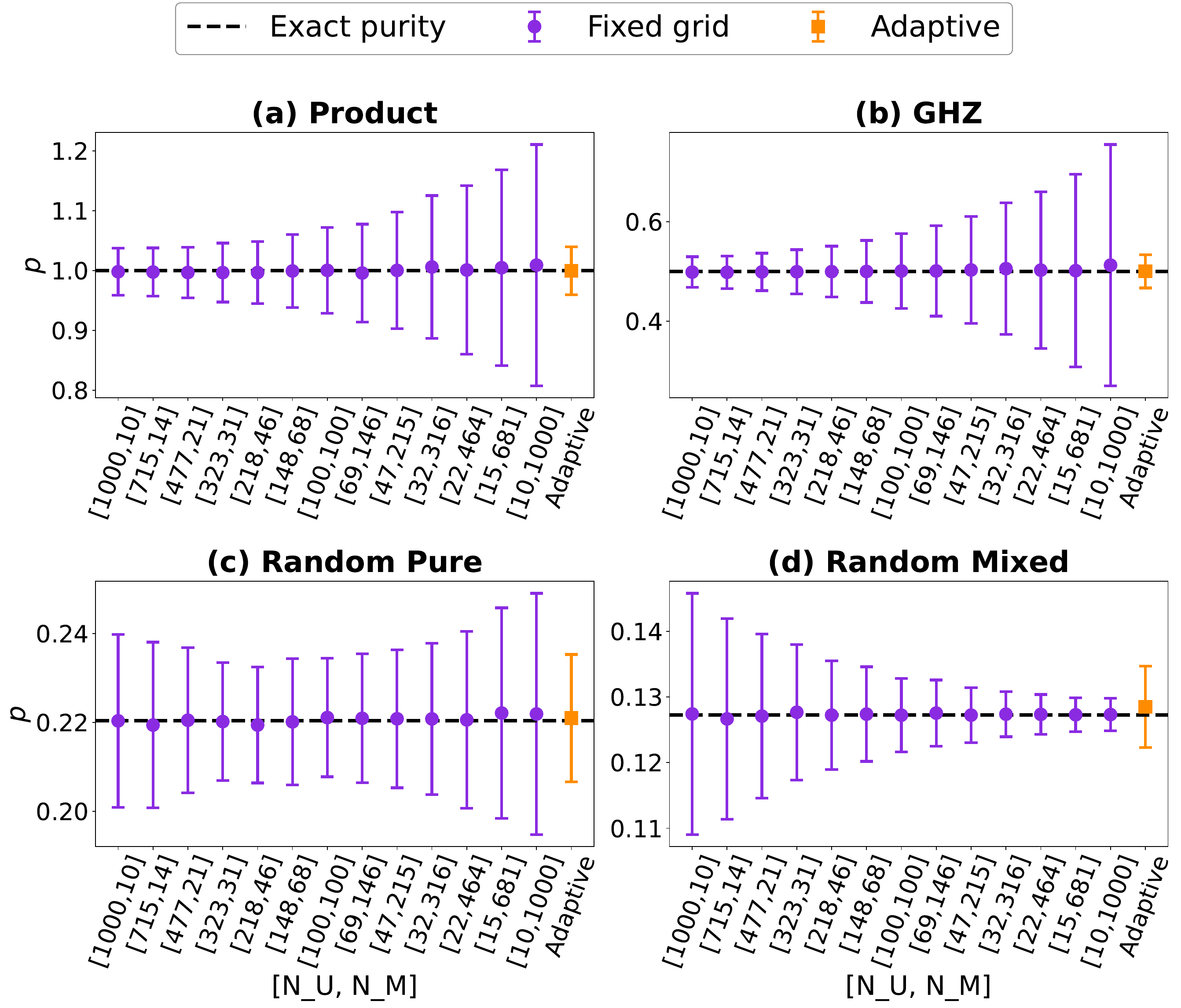}
    \caption{Comparison between the adaptive protocol and a grid search over fixed $(N_U,N_M)$ parameter choices. For each grid point and for the adaptive protocol, we perform 500 independent estimation trials on the half-system partition of several 6-qubit states: (a) product state, (b) GHZ state, (c) Haar-random pure state, and (d) random mixed state. The plotted points show the mean over the 500 trials, with error bars indicating one standard deviation. The grid consists of 13 parameter sets with $N_M$ generated through logarithmic spacing. For each $N_M$, $N_U$ is chosen as the smallest integer such that $N_U N_M \geq 10000$. The adaptive protocol is run with the same total measurement budget, $N_{\mathrm{tot}}=10000$, ensuring that the comparison is not biased in favor of the adaptive method. Across all states considered, the adaptive protocol achieves performance comparable to the best grid points: measured by the standard deviation of the estimates, it outperforms 12 of the 13 grid points for the product state, 11 of the 13 grid points for the GHZ state, 9 of the 13 grid points for the Haar-random pure state, and 6 of the 13 grid points for the random mixed state.
}
    \label{fig:Comparison with Grid}
\end{figure}

To better demonstrate the power of the adaptive method, we generate a grid of $(N_U,N_M)$ pairs, run the randomized protocol at each grid point many times, and compare it with the adaptive method. We repeat this comparison for several classes of states with different purities: product, GHZ, random pure, and random mixed states. The grid experiment and the adaptive trials are given the same total budget $N_{\mathrm{tot}}=10000$. As shown in Fig.~\ref{fig:Comparison with Grid}, we run each grid point and the adaptive method 500 times. We plot the mean of the purity estimate $\widehat{p}$ over the 500 trials together with its standard deviation, and the exact purity is shown as a horizontal line. We see that for the product and GHZ states, the adaptive method performs comparably to the best grid point, while for the random pure and random mixed states its standard deviation remains within a factor of about $1.2$ and $2.5$ of the best grid point, respectively. The grid analysis also confirms that the randomized protocol performs better with a larger $N_U/N_M$ ratio for highly pure states, and with a smaller $N_U/N_M$ ratio for highly mixed states.

\begin{figure*}
    \centering
    \includegraphics[width=1\linewidth]{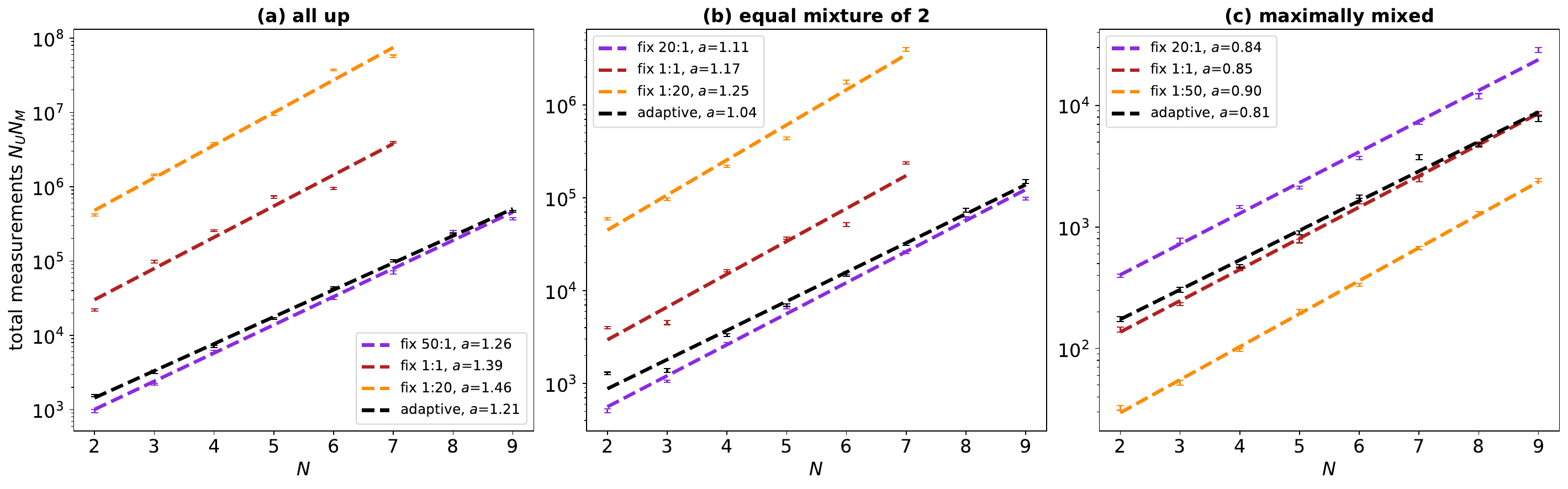}
    \caption{Scaling of the experimental resources required to achieve an estimation accuracy $\abs{\widehat{p} - p}\leq 0.1$ as a function of system size $N$ (from $N=2$ to $N=9$), where the measured subsystem is chosen to be the full system ($N_A = N$). For each state, we compare three fixed-ratio protocols, corresponding to a large $N_U/N_M$ ratio, the balanced case $N_U/N_M=1$, and a small $N_U/N_M$ ratio; the ratios shown are $N_U/N_M = 50, 1, 1/20$ in panel (a), $20, 1, 1/20$ in panel (b), and $20, 1, 1/50$ in panel (c). In each fixed-ratio protocol, $(N_U,N_M)$ is determined by the prescribed ratio and allocated total resource $N_U N_M$. Along each fixed-ratio curve, all data points therefore share the same value of $N_U/N_M$, even though different system sizes require different total resources to reach the target accuracy. The adaptive protocol is shown alongside these fixed-ratio protocols. Each curve is fitted to an exponential form $2^{aN+b}$, with the fitted exponent $a$ reported in the legend. Each plotted data point is represented by an error bar that bounds the minimum resource required to reach the target accuracy. Specifically, the resource value at the top of the error bar is sufficient for the interval defined by the mean estimate over 100 trials plus or minus one standard deviation to lie entirely within $p\pm 0.1$. In contrast, the resource value at the bottom of the error bar is insufficient to satisfy this criterion. Thus, each error bar provides an upper and lower bound on the minimum required resource. For the exponential fits, we use the mean value of each error bar. (a) All-spin-up state $\ket{0}^{\otimes N}$, with purity $1$. (b) Equal mixture $\frac{1}{2}\left[(\ket{0}\bra{0})^{\otimes N}+(\ket{1}\bra{1})^{\otimes N}\right]$, with purity $0.5$. (c) Maximally mixed state, with purity $1/2^N$. These states of size $N$ can be seen as the reduced states of the product, GHZ, and maximally mixed states of size $2N$, respectively.}
    \label{fig:Resource Scaling}
\end{figure*}

Note that Fig.~\ref{fig:Comparison with Grid} only shows the performance of the adaptive protocol at a single system size, $N=6$, where it is not yet able to outperform the best fixed-ratio protocol for the random pure and random mixed states. To better study the effectiveness of the adaptive protocol, we need to determine how its required experimental resource scales with system size and benchmark it against the fixed-ratio protocols.

To this end, we investigate how the total experimental resource $N_U N_M$ scales with system size $N$ to reach a fixed accuracy $\abs{\widehat{p} - p} \leq 0.1$. Here we take $N_A = N$ as the measured subsystem, rather than a half partition of the system as in the previous plots, because this allows us to obtain data points at odd system sizes. The three states we study are the all-spin-up state $\ket{0}^{\otimes N}$ with purity $1$, the equal mixture $\frac{1}{2}\left[(\ket{0}\bra{0})^{\otimes N}+(\ket{1}\bra{1})^{\otimes N}\right]$ with purity $0.5$, and the maximally mixed state with purity $1/2^N$. These states cover the maximal and minimal purity allowed at each system size, as well as an intermediate value. They can also be viewed as reduced states of the states tested in the previous plots: the half-partition reduced states of the product state and the maximally mixed state remain product and maximally mixed states, respectively, while the half-partition reduced state of the GHZ state is the equal mixture state.

For each state, we compare the adaptive protocol against three fixed-ratio protocols with large, balanced ($N_U/N_M = 1$), and small $N_U/N_M$ ratios. The fixed-ratio curves use standard fixed-parameter estimation at each data point, and along each fixed-ratio curve, the $N_U/N_M$ ratio is kept constant. Given this constraint, the total resource $N_U N_M$ is tuned at every system size to find the minimal resource. The resulting scaling curves are fitted to $2^{aN+b}$, with the exponents $a$ reported in the legend of Fig.~\ref{fig:Resource Scaling}. We expect this scaling behavior on general grounds, since any protocol that estimates the purity of an $N$-qubit state to constant additive error using single-copy measurements requires $\Omega(2^{N/2})$ samples~\cite{Gong2026}.

As shown in Fig.~\ref{fig:Resource Scaling}, suboptimal $N_U/N_M$ ratios require significantly more, sometimes exponentially more, resources. In the all-spin-up state case, the worst $N_U/N_M$ ratio in the plot ($N_U/N_M = 1/20$) requires $2^{8}$--$2^{10}$ times more resource than the adaptive protocol across system sizes $N = 3$--$7$. For the equal mixture case, the worst ratio scales with both a higher exponent and a larger constant factor. The figure also shows that the adaptive protocol exhibits the slowest growth for all three states tested. 

\section{Conclusion}

We have presented an adaptive protocol for estimating the second-order R\'enyi entropy, an important quantity for studying and characterizing quantum many-body systems. Our protocol extends the randomized measurement protocol framework of Ref.~\cite{brydges2019probing}. It begins with a small amount of data and iteratively decides whether to increase $N_U$ or $N_M$ until the total measurement budget is exhausted. By running the protocol many times and analyzing the distribution of the final allocation $(N_U, N_M)$, and by benchmarking it against a grid of fixed-parameter estimators, we find that, for a given system size, the adaptive protocol reliably adapts to a favorable parameter regime.

We also investigated how the minimal number of samples required to reach a fixed additive error in purity estimation scales with system size, for several representative states covering minimal, intermediate, and maximal purities. For each system size, we plotted the minimum resource required for purity estimation to be within some additive error for the adaptive protocol, as well as three fixed-ratio curves, covering small, balanced, and large $N_U/N_M$ ratios. We find that the adaptive protocol exhibits a smaller scaling exponent than all three fixed-ratio curves tested, for all representative states tested.

Our work thus extends the randomized measurement protocol to the state-blind setting at no additional experimental cost. Identifying the tasks in which adaptivity reduces sample complexity is an important open problem in quantum learning. In particular, whether adaptivity helps may depend on the metric under consideration~\cite{Chen2023}. Here we provide an example in which adaptivity helps in two ways. First, the best parameter ratio depends on the initial state being estimated and cannot be chosen \textit{a priori}; the adaptive method finds this favorable regime on its own and, at fixed system size, performs comparably to the best fixed-parameter estimator for states with high and intermediate purity, without prior knowledge of the state. Second, over the range of system sizes studied, the adaptive protocol exhibits a smaller scaling exponent than each of the fixed-ratio protocols tested for states with maximal, intermediate, and minimal purities, so its advantage over any fixed-ratio choice grows with system size. The scaling comparisons are made against fixed-ratio protocols; we do not rule out that a non-adaptive protocol whose ratio is optimally tuned at every system size could achieve the same scaling.

Finally, the adaptive strategy developed here could potentially be extended beyond the second-order R\'enyi entropy and purity. For instance, 
estimating the $n^{\text{th}}$-order R\'enyi entropy requires estimating $\Tr(\rho^n)$, which, as a generalization of the second-order case, is equivalent to estimating the expectation value of the $n$-copy cyclic permutation operator. Directly measuring this operator requires coherent manipulation of $n$ copies of the quantum state. For $n > 2$ the cyclic permutation operator is no longer Hermitian, and the non-Hermiticity must be bypassed using techniques such as the Hadamard test~\cite{Ekert2002}. To the best of our knowledge, there has been no experimental demonstration of a multi-copy coherent generalized SWAP test for estimating $\Tr(\rho^n)$ with $n > 2$. Single-copy randomized measurement estimators of $\Tr(\rho^n)$ have, however, been demonstrated experimentally~\cite{Rath2023}, even though the estimator involves all ordered $n$-tuples of the $N_{\mathrm{tot}}$ shadows, so that the number of terms grows as $N_{\mathrm{tot}}!/(N_{\mathrm{tot}} - n)!$. The von Neumann entropy can be viewed as an analytic continuation of the R\'enyi entropies, and estimating it in this way would in principle require the R\'enyi entropies of all orders. Recent work using a stabilized analytic continuation (SAC) approach reduces this requirement to a finite number of orders~\cite{Vijay2026}. Another possible generalization is $\Tr(O\rho^2)$ with $O$ a generic observable, which reduces to the purity for $O = I$. This problem admits a randomized protocol with an explicit $(N_U, N_M)$ construction~\cite{du2026optimal}, to which our adaptive protocol could potentially be applied. \\

\begin{acknowledgments}
    We acknowledge financial support from the Natural Sciences and Engineering Research Council of Canada (NSERC) Discovery program (RGPIN-2025-06496), the University of Waterloo, and The Strategic Science Fund of the Government of Canada. This work made use of ``Symmetry'', the high-performance computing system at Perimeter Institute for Theoretical Physics.
    Research at the Perimeter Institute is supported in part by the Government of Canada through the Department of Innovation, Science and Economic Development Canada and by the Province of Ontario through the Ministry of Economic Development, Job Creation and Trade.
\end{acknowledgments}


\bibliography{Refs}

@article{elben2019statistical,
  author  = {Elben, Andreas and Vermersch, Beno{\^i}t and Roos, Christian F. and Zoller, Peter},
  title   = {Statistical Correlations between Locally Randomized Measurements: A Toolbox for Probing Entanglement in Many-Body Quantum States},
  journal = {Phys. Rev. A},
  volume  = {99},
  pages   = {052323},
  year    = {2019},
  doi     = {10.1103/PhysRevA.99.052323}
}

@article{du2026optimal,
  author  = {Du, Zhenyu and Tang, Yifan and Elben, Andreas and Roth, Ingo and Eisert, Jens and Liu, Zhenhuan},
  title   = {Optimal Randomized Measurements for a Family of Nonlinear Quantum Properties},
  journal = {PRX Quantum},
  volume  = {7},
  pages   = {010360},
  year    = {2026},
  doi     = {10.1103/4rkx-xkwy}
}

@article{abanin2012quantum,
  author  = {Abanin, Dmitry A. and Demler, Eugene},
  title   = {Measuring Entanglement Entropy of a Generic Many-Body System with a Quantum Switch},
  journal = {Phys. Rev. Lett.},
  volume  = {109},
  pages   = {020504},
  year    = {2012},
  doi     = {10.1103/PhysRevLett.109.020504}
}

@article{daley2012entanglement,
  author  = {Daley, A. J. and Pichler, H. and Schachenmayer, J. and Zoller, P.},
  title   = {Measuring Entanglement Growth in Quench Dynamics of Bosons in an Optical Lattice},
  journal = {Phys. Rev. Lett.},
  volume  = {109},
  pages   = {020505},
  year    = {2012},
  doi     = {10.1103/PhysRevLett.109.020505}
}

@article{islam2015entanglement,
  author  = {Islam, Rajibul and Ma, Ruichao and Preiss, Philipp M. and Tai, M. Eric and Lukin, Alexander and Rispoli, Matthew and Greiner, Markus},
  title   = {Measuring Entanglement Entropy in a Quantum Many-Body System},
  journal = {Nature},
  volume  = {528},
  pages   = {77--83},
  year    = {2015},
  doi     = {10.1038/nature15750}
}

@article{vermersch2018unitary,
  author  = {Vermersch, Beno{\^i}t and Elben, Andreas and Dalmonte, Marcello and Cirac, J. Ignacio and Zoller, Peter},
  title   = {Unitary $n$-designs via random quenches in atomic {Hubbard} and spin models: Application to the measurement of R\'enyi entropies},
  journal = {Phys. Rev. A},
  volume  = {97},
  pages   = {023604},
  year    = {2018},
  doi     = {10.1103/PhysRevA.97.023604}
}

@article{brydges2019probing,
  author  = {Brydges, Tiff and Elben, Andreas and Jurcevic, Petar and Vermersch, Beno{\^i}t and Maier, Christine and Lanyon, Benjamin P. and Zoller, Peter and Blatt, Rainer and Roos, Christian F.},
  title   = {Probing R\'enyi Entanglement Entropy via Randomized Measurements},
  journal = {Science},
  volume  = {364},
  pages   = {260--263},
  year    = {2019},
  doi     = {10.1126/science.aau4963}
}

@article{zhou2024hybrid,
  author  = {Zhou, You and Liu, Zhenhuan},
  title   = {A Hybrid Framework for Estimating Nonlinear Functions of Quantum States},
  journal = {npj Quantum Inf.},
  volume  = {10},
  pages   = {62},
  year    = {2024},
  doi     = {10.1038/s41534-024-00846-5}
}

@article{elben2018renyi,
  author  = {Elben, Andreas and Vermersch, Beno{\^i}t and Dalmonte, Marcello and Cirac, J. Ignacio and Zoller, Peter},
  title   = {R\'enyi Entropies from Random Quenches in Atomic {Hubbard} and Spin Models},
  journal = {Phys. Rev. Lett.},
  volume  = {120},
  pages   = {050406},
  year    = {2018},
  doi     = {10.1103/PhysRevLett.120.050406}
}

@misc{plodzien2025scrambling,
  author       = {P{\l}odzie{\'n}, Marcin},
  title        = {Lecture Notes on Information Scrambling, Quantum Chaos, and {Haar}-Random States},
  year         = {2025},
  eprint       = {2511.14397},
  archivePrefix= {arXiv},
  primaryClass = {quant-ph},
  doi          = {10.48550/arXiv.2511.14397}
}

@article{huang2020predicting,
  author  = {Huang, Hsin-Yuan and Kueng, Richard and Preskill, John},
  title   = {Predicting Many Properties of a Quantum System from Very Few Measurements},
  journal = {Nat. Phys.},
  volume  = {16},
  pages   = {1050--1057},
  year    = {2020},
  doi     = {10.1038/s41567-020-0932-7}
}

@article{Fan,
  author  = {Fan, Ruihua and Bao, Yimu and Altman, Ehud and Vishwanath, Ashvin},
  title   = {Diagnostics of Mixed-State Topological Order and Breakdown of Quantum Memory},
  journal = {PRX Quantum},
  volume  = {5},
  pages   = {020343},
  year    = {2024},
  doi     = {10.1103/PRXQuantum.5.020343}
}

@article{Grusdt,
  author  = {Grusdt, F.},
  title   = {Topological order of mixed states in correlated quantum many-body systems},
  journal = {Phys. Rev. B},
  volume  = {95},
  pages   = {075106},
  year    = {2017},
  doi     = {10.1103/PhysRevB.95.075106}
}

@article{Tarun,
  author  = {Wang, Ting-Tung and Song, Menghan and Meng, Zi Yang and Grover, Tarun},
  title   = {Analog of Topological Entanglement Entropy for Mixed States},
  journal = {PRX Quantum},
  volume  = {6},
  pages   = {010358},
  year    = {2025},
  doi     = {10.1103/PRXQuantum.6.010358}
}

@article{Choi,
  author  = {Choi, Soonwon and Bao, Yimu and Qi, Xiao-Liang and Altman, Ehud},
  title   = {Quantum Error Correction in Scrambling Dynamics and Measurement-Induced Phase Transition},
  journal = {Phys. Rev. Lett.},
  volume  = {125},
  pages   = {030505},
  year    = {2020},
  doi     = {10.1103/PhysRevLett.125.030505}
}

@article{Gullans,
  author  = {Gullans, Michael J. and Huse, David A.},
  title   = {Dynamical Purification Phase Transition Induced by Quantum Measurements},
  journal = {Phys. Rev. X},
  volume  = {10},
  pages   = {041020},
  year    = {2020},
  doi     = {10.1103/PhysRevX.10.041020}
}

@article{Chen,
  author  = {Chen, Yu-Hsueh and Grover, Tarun},
  title   = {Separability Transitions in Topological States Induced by Local Decoherence},
  journal = {Phys. Rev. Lett.},
  volume  = {132},
  pages   = {170602},
  year    = {2024},
  doi     = {10.1103/PhysRevLett.132.170602}
}

@article{KP,
  author  = {Kitaev, Alexei and Preskill, John},
  title   = {Topological Entanglement Entropy},
  journal = {Phys. Rev. Lett.},
  volume  = {96},
  pages   = {110404},
  year    = {2006},
  doi     = {10.1103/PhysRevLett.96.110404}
}

@article{LW,
  author  = {Levin, Michael and Wen, Xiao-Gang},
  title   = {Detecting Topological Order in a Ground State Wave Function},
  journal = {Phys. Rev. Lett.},
  volume  = {96},
  pages   = {110405},
  year    = {2006},
  doi     = {10.1103/PhysRevLett.96.110405}
}

@article{CC,
  author  = {Calabrese, Pasquale and Cardy, John},
  title   = {Entanglement entropy and quantum field theory},
  journal = {J. Stat. Mech.},
  volume  = {2004},
  pages   = {P06002},
  year    = {2004},
  doi     = {10.1088/1742-5468/2004/06/P06002}
}

@article{FM,
  author  = {Fradkin, Eduardo and Moore, Joel E.},
  title   = {Entanglement Entropy of {2D} Conformal Quantum Critical Points: Hearing the Shape of a Quantum Drum},
  journal = {Phys. Rev. Lett.},
  volume  = {97},
  pages   = {050404},
  year    = {2006},
  doi     = {10.1103/PhysRevLett.97.050404}
}

@article{MIPT1,
  author  = {Skinner, Brian and Ruhman, Jonathan and Nahum, Adam},
  title   = {Measurement-Induced Phase Transitions in the Dynamics of Entanglement},
  journal = {Phys. Rev. X},
  volume  = {9},
  pages   = {031009},
  year    = {2019},
  doi     = {10.1103/PhysRevX.9.031009}
}

@article{MIPT2,
  author  = {Li, Yaodong and Chen, Xiao and Fisher, Matthew P. A.},
  title   = {Measurement-driven entanglement transition in hybrid quantum circuits},
  journal = {Phys. Rev. B},
  volume  = {100},
  pages   = {134306},
  year    = {2019},
  doi     = {10.1103/PhysRevB.100.134306}
}

@article{Negari,
  author  = {Negari, Amir-Reza and Ellison, Tyler D. and Hsieh, Timothy H.},
  title   = {Spacetime {Markov} length: a diagnostic for fault tolerance via mixed-state phases},
  journal = {Phys. Rev. B},
  volume  = {114},
  pages   = {045130},
  year    = {2026},
  doi     = {10.1103/61wf-zx39}
}

@article{Sang,
  author  = {Sang, Shengqi and Hsieh, Timothy H.},
  title   = {Stability of Mixed-State Quantum Phases via Finite {Markov} Length},
  journal = {Phys. Rev. Lett.},
  volume  = {134},
  pages   = {070403},
  year    = {2025},
  doi     = {10.1103/PhysRevLett.134.070403}
}

@article{Dankert2009ExactApproximateUnitary2Designs,
  author  = {Dankert, Christoph and Cleve, Richard and Emerson, Joseph and Livine, Etera},
  title   = {Exact and Approximate Unitary 2-Designs: Constructions and Applications},
  journal = {Phys. Rev. A},
  volume  = {80},
  pages   = {012304},
  year    = {2009},
  doi     = {10.1103/PhysRevA.80.012304}
}

@inproceedings{Chen2022,
  author    = {Chen, Sitan and Cotler, Jordan and Huang, Hsin-Yuan and Li, Jerry},
  title     = {Exponential Separations Between Learning With and Without Quantum Memory},
  booktitle = {2021 {IEEE} 62nd Annual Symposium on Foundations of Computer Science (FOCS)},
  pages     = {574--585},
  year      = {2022},
  publisher = {IEEE},
  doi       = {10.1109/FOCS52979.2021.00063}
}

@inproceedings{Chen2023,
  author    = {Chen, Sitan and Huang, Brice and Li, Jerry and Liu, Allen and Sellke, Mark},
  title     = {When Does Adaptivity Help for Quantum State Learning?},
  booktitle = {2023 {IEEE} 64th Annual Symposium on Foundations of Computer Science (FOCS)},
  pages     = {391--404},
  year      = {2023},
  publisher = {IEEE},
  doi       = {10.1109/FOCS57990.2023.00029}
}

@article{Ekert2002,
  author  = {Ekert, Artur K. and Moura Alves, Carolina and Oi, Daniel K. L. and Horodecki, Micha{\l} and Horodecki, Pawe{\l} and Kwek, L. C.},
  title   = {Direct Estimations of Linear and Nonlinear Functionals of a Quantum State},
  journal = {Phys. Rev. Lett.},
  volume  = {88},
  pages   = {217901},
  year    = {2002},
  doi     = {10.1103/PhysRevLett.88.217901}
}

@article{Rath2023,
  author  = {Rath, Aniket and Vitale, Vittorio and Murciano, Sara and Votto, Matteo and Dubail, J{\'e}r{\^o}me and Kueng, Richard and Branciard, Cyril and Calabrese, Pasquale and Vermersch, Beno{\^i}t},
  title   = {Entanglement Barrier and Its Symmetry Resolution: Theory and Experimental Observation},
  journal = {PRX Quantum},
  volume  = {4},
  pages   = {010318},
  year    = {2023},
  doi     = {10.1103/PRXQuantum.4.010318}
}

@article{Ransford2026,
  author  = {Ransford, Anthony and Allman, M. S. and Arkinstall, Jake and Campora, III, J. P. and Cooper, Samuel F. and Delaney, Robert D. and Dreiling, Joan M. and Estey, Brian and Figgatt, Caroline and Hall, Alex and others},
  title   = {A 98-qubit trapped-ion quantum computer with all-to-all connectivity},
  journal = {Nature},
  volume  = {655},
  pages   = {81--86},
  year    = {2026},
  doi     = {10.1038/s41586-026-10676-4}
}

@article{Bruzewicz2019,
  author  = {Bruzewicz, Colin D. and Chiaverini, John and McConnell, Robert and Sage, Jeremy M.},
  title   = {Trapped-Ion Quantum Computing: Progress and Challenges},
  journal = {Appl. Phys. Rev.},
  volume  = {6},
  pages   = {021314},
  year    = {2019},
  doi     = {10.1063/1.5088164}
}

@article{Saffman2016,
  author  = {Saffman, M.},
  title   = {Quantum Computing with Atomic Qubits and Rydberg Interactions: Progress and Challenges},
  journal = {J. Phys. B},
  volume  = {49},
  pages   = {202001},
  year    = {2016},
  doi     = {10.1088/0953-4075/49/20/202001}
}

@article{Evered2023,
  author  = {Evered, Simon J. and Bluvstein, Dolev and Kalinowski, Marcin and Ebadi, Sepehr and Manovitz, Tom and Zhou, Hengyun and Li, Sophie H. and Geim, Alexandra A. and Wang, Tout T. and Maskara, Nishad and Levine, Harry and Semeghini, Giulia and Greiner, Markus and Vuleti{\'c}, Vladan and Lukin, Mikhail D.},
  title   = {High-fidelity parallel entangling gates on a neutral-atom quantum computer},
  journal = {Nature},
  volume  = {622},
  pages   = {268--272},
  year    = {2023},
  doi     = {10.1038/s41586-023-06481-y}
}

@article{Vijay2026,
  author    = {Vijay, Akash and Raj, Ayush and Kudler-Flam, Jonah and Vermersch, Beno{\^i}t and Elben, Andreas and Nie, Laimei},
  title     = {Analytically Continuing the Randomized Measurement Toolbox},
  journal   = {Phys. Rev. Lett.},
  volume    = {137},
  issue     = {10},
  pages     = {100202},
  numpages  = {7},
  year      = {2026},
  month     = {Sep},
  publisher = {American Physical Society},
  doi       = {10.1103/837t-68bf},
  url       = {https://link.aps.org/doi/10.1103/837t-68bf}
}

@article{Elben2023Toolbox,
  author  = {Elben, Andreas and Flammia, Steven T. and Huang, Hsin-Yuan and Kueng, Richard and Preskill, John and Vermersch, Beno{\^i}t and Zoller, Peter},
  title   = {The Randomized Measurement Toolbox},
  journal = {Nat. Rev. Phys.},
  volume  = {5},
  pages   = {9--24},
  year    = {2023},
  doi     = {10.1038/s42254-022-00535-2}
}

@book{blitzstein_hwang_2019_probability,
    author    = {Blitzstein, Joseph K. and Hwang, Jessica},
    title     = {Introduction to Probability},
    edition   = {2},
    publisher = {Chapman and Hall/CRC},
    address   = {Boca Raton, FL},
    year      = {2019},
    chapter   = {9},
    pages     = {415--456},
    isbn      = {978-1-1383-6991-7}
}

@article{Hoeffding1948,
  author  = {Hoeffding, Wassily},
  title   = {A class of statistics with asymptotically normal distribution},
  journal = {Ann. Math. Statist.},
  volume  = {19},
  issue   = {3},
  pages   = {293--325},
  year    = {1948},
  publisher = {Institute of Mathematical Statistics},
  doi     = {10.1214/aoms/1177730196}
}

@article{Hayden2006,
  title = {Aspects of generic entanglement},
  author = {Hayden, Patrick and Leung, Debbie W. and Winter, Andreas},
  journal = {Commun. Math. Phys.},
  volume = {265},
  number = {1},
  pages = {95--117},
  year = {2006},
  doi = {10.1007/s00220-006-1535-6},
}

@inbook{Lipschutz2011Ch3,
  author    = {Lipschutz, Seymour and Schiller, John J.},
  title     = {Basic Probability},
  booktitle = {Schaum's Outline of Introduction to Probability and Statistics},
  publisher = {McGraw-Hill},
  address   = {New York},
  year      = {2011},
  chapter   = {3},
  pages     = {78},
  isbn      = {978-0-07-176249-6}
}

@article{Gong2026,
  author  = {Gong, Weiyuan and Haferkamp, Jonas and Ye, Qi and Zhang, Zhihan},
  title   = {Sample complexity of purity and inner product estimation},
  journal = {Phys. Rev. Res.},
  volume  = {8},
  pages   = {043077},
  year    = {2026},
  doi     = {10.1103/kh2j-jkn6}
}

@article{Johansson2012QuTiP,
  author  = {Johansson, J. R. and Nation, P. D. and Nori, Franco},
  title   = {{QuTiP}: An open-source {Python} framework for the dynamics of open quantum systems},
  journal = {Comput. Phys. Commun.},
  volume  = {183},
  pages   = {1760--1772},
  year    = {2012},
  doi     = {10.1016/j.cpc.2012.02.021}
}

\newpage

\appendix

\section{Unbiased Estimator of Purity} \label{Unbiased Estimator for Purity}

We split the proof into two parts. First, we show that the expression in terms of the Hamming kernel in Eq.~\eqref{eq: Purity Estimator Kernel Form} is an unbiased estimator for the purity. Second, we show that this Hamming kernel form can be written in the count-space form of Eq.~\eqref{eq:estimator for TrXU count space}.

\begin{lemma}\label{lem:shot-expectation-Hamming-kernel}
    For two distinct samples $i$ and $j$ ($i\neq j$), 
    \begin{equation}
        \Eshot{W_{s_i s_j}}{U} = \vb{p}_U^{\top} W \vb{p}_U.
    \end{equation}
\end{lemma}

\begin{proof}
    \[
    \begin{split}
        \Eshot{W_{s_i s_j}}{U} 
        &= \sum_{s,\, s'} \Pr\left(s_i = s,\, s_j = s' \,\middle|\, U\right)\, W_{s s'} \\
        &= \sum_{s,\, s'} P_U(s) P_U(s') W_{s s'} \\
        &= \vb{p}_U^{\top} W \vb{p}_U
    \end{split}
    \]
    where $\Pr\left(s_i = s,\, s_j = s' \,\middle|\, U\right)$ denotes the probability that we get string $s$ in the $i^{\text{th}}$ shot and string $s'$ in the $j^{\text{th}}$ shot.
\end{proof}

\begin{proposition}\label{prp:unbiased estimator for TrXU}
    $\widehat{\mathcal{X}_U}$ is an unbiased estimator for $\mathcal{X}_U$ in the sense that
\begin{equation}
    \Eshot{\widehat{\mathcal{X}_U}}{U}  = \mathcal{X}_U.
\end{equation}    
\end{proposition}

\begin{proof}
    We use the definition of $\widehat{\mathcal{X}_U}$ in Eq.~\eqref{eq: Purity Estimator Kernel Form} and \cref{lem:shot-expectation-Hamming-kernel}
    \[
    \begin{split}
        \Eshot{\widehat{\mathcal{X}_U}}{U}  
        &= \frac{2^{N_A}}{(N_M)^{\underline{2}}} \sum_{i,j \,|\, i \neq j} \Eshot{W_{s_i s_j}}{U} \\
        &= \frac{2^{N_A}}{(N_M)^{\underline{2}}} \sum_{i,j \,|\, i \neq j} \vb{p}_U^{\top} W \vb{p}_U \\
        &= 2^{N_A} \vb{p}_U^{\top} W \vb{p}_U
    \end{split}
    \]
    where we used that the number of ordered 2-tuples chosen from $N_M$ objects is exactly $(N_M)^{\underline{2}}$. We then expand $\mathcal{X}_U$:
    \begin{equation}
        \label{eq: XU in terms of probability}
        \begin{split}
        \mathcal{X}_U
        &= \Tr[U^{\dagger\otimes 2} X U^{\otimes 2} \rho_A^{\otimes 2}] \\
        &= 2^{N_A} \sum_{s,s'} W_{ss'}
        \Tr[\ket{s}\bra{s} U\rho_A U^\dagger] \\
        & \qquad \times
        \Tr[\ket{s'}\bra{s'} U\rho_A U^\dagger] \\
        &= 2^{N_A} \sum_{s,s'} W_{ss'} P_U(s) P_U(s')\\
        &= 2^{N_A} \vb{p}_U^{\top} W \vb{p}_U 
        \end{split}
    \end{equation}
    Hence, $ \Eshot{\widehat{\mathcal{X}_U}}{U}  = \mathcal{X}_U$.

\end{proof}

\begin{theorem}
    \label{thm:unbiased estimator for purity}
    $\widehat{\mathcal{X}_U}$ is an unbiased estimator for purity $p(\rho,A)$ in the sense that
    \[
        \E[U]{\Eshot{\widehat{\mathcal{X}_U}}{U}} = p(\rho,A).
    \]
\end{theorem}

\begin{proof}
    It remains to show that $\E[U]{\mathcal{X}_U} = p(\rho,A)$.
    \[
    \begin{split}
        \E[U]{\mathcal{X}_U} &= \E[U]{\Tr[X_U \rho_A^{\otimes 2}]} \\
        &= \Tr[\E[U]{X_U}\rho_A^{\otimes 2}] \\
        &= \Tr[\operatorname{SWAP} \rho_A^{\otimes 2}]\\
        &= p(\rho,A)
    \end{split}
    \]
    Then by \cref{prp:unbiased estimator for TrXU},
    \[
    \E[U]{\Eshot{\widehat{\mathcal{X}_U}}{U}} = \E[U]{\mathcal{X}_U} = p(\rho,A)
    \]
\end{proof}

\begin{theorem}
    \label{thm:unbiased estimator for purity count space}
    The unbiased estimator for purity $\widehat{\mathcal{X}_U}$ can be expressed in count space as shown in Eq.~\eqref{eq:estimator for TrXU count space}, i.e.
    \[
    \widehat{\mathcal{X}_U} = \frac{2^{N_A}}{(N_M)^{\underline{2}}} \left(\vb{c}_U^{\top} W \vb{c}_U - N_M\right).
    \]
\end{theorem}

\begin{proof}
    The definition of $\widehat{\mathcal{X}_U}$ in Eq.~\eqref{eq: Purity Estimator Kernel Form} is a sum over all ordered pairs of distinct shots. Note that two distinct shots $i \neq j$ may still give the same outcome string, and such pairs remain included in the sum. The count space form rewrites this shot sum using the outcome counts. First note that
    \[
    \sum_{i,j} W_{s_i s_j} = \sum_{s,s'} c_U(s)\, c_U(s')\, W_{s s'} = \vb{c}_U^{\top} W \vb{c}_U
    \]
    where we group the shots according to the outcome strings obtained at these shots, and used that the kernel value depends only on the pair of strings appearing in the index. The unrestricted sum $\sum_{i,j}$ differs from $\sum_{i,j | i \neq j}$ only by the same-shot diagonal terms with $i = j$: there are $N_M$ such terms, one for each shot, and each of them contributes $W_{s_i s_i} = 1$. The count space expression therefore removes only these $N_M$ same-shot contributions, while pairs of distinct shots with the same outcome string remain included. Hence,
    \begin{equation}
    \label{sum hamming kernel distinct shot}
    \sum_{i,j \,|\, i \neq j} W_{s_i s_j} = \vb{c}_U^{\top} W \vb{c}_U - N_M
    \end{equation}
    Then, substituting Eq.~\eqref{sum hamming kernel distinct shot} into Eq.~\eqref{eq: Purity Estimator Kernel Form}, we get the count space form in Eq.~\eqref{eq:estimator for TrXU count space}.
\end{proof}

\section{Variance Analysis of Randomized Protocol} \label{app:Variance Analysis of Randomized Protocol}

\begin{lemma}[Adam's Law]\label{lem: adams law}
    For random variables $X$ and $Y$,
    \[
    \E{\Ec{Y}{X}} = \E{Y}
    \]
\end{lemma}

\begin{proof}
    Intuitively, averaging the conditional expectation of random variable $Y$ over all conditions should just give us the expectation of $Y$. For a complete proof, see Theorem 9.3.7 in Ref.~\cite{blitzstein_hwang_2019_probability}.
\end{proof}

\begin{theorem}[Eve's Law]\label{thm: eves law}
    For random variables $X$ and $Y$, 
    \[
    \V{Y} = \E{\Vc{Y}{X}} + \V{\Ec{Y}{X}}
    \]
    This is also called the law of total variance.
\end{theorem}

\begin{proof}
    We expand the two terms on the right-hand side in turn.

    By Lemma~\ref{lem: adams law}, we have $\E{\Ec{Y^2}{X}} = \E{Y^2}$, so
    \[
    \begin{split}
        \E{\Vc{Y}{X}} &= \E{ \Ec{Y^2}{X} -  \Ec{Y}{X}^2 }\\
        &= \E{Y^2} - \E{ \Ec{Y}{X}^2 }
    \end{split}
    \]
    Still by Lemma~\ref{lem: adams law}, we have
    \[
    \begin{split}
        \V{\Ec{Y}{X}} &= \E{ \Ec{Y}{X}^2 } - \E{ \Ec{Y}{X} }^2 \\
        &= \E{ \Ec{Y}{X}^2 } - \E{Y}^2
    \end{split}
    \]
    Then,
    \[
    \begin{split}
        \E{\Vc{Y}{X}} + \V{\Ec{Y}{X}} &= \E{Y^2} - \E{Y}^2 \\
        &= \V{Y}
    \end{split}
    \]

\end{proof}

\begin{corollary}
    The variance of purity estimator $\widehat{p} \equiv \frac{1}{N_U} \sum_{r=1}^{N_U} \widehat{\mathcal{X}_{U_r}}$ can be expressed as in Eq.~\eqref{eq:total variance}, i.e.
    \[
    \V{\widehat{p}} = \frac{V_U + V_{\mathrm{shot}}(N_M)}{N_U}
    \]
    where $V_U = \V[U]{\mathcal{X}_U}$ and $V_{\mathrm{shot}}(N_M) = \E[U]{\Vshot{\widehat{\mathcal{X}_{U}}}{U}}$
\end{corollary}

\begin{proof}
    We first distribute the variance inside the sum using that for a random variable $X$, $\V[]{cX} = c^2 \V[]{X}$ and that for independent random variables $X$ and $Y$, $\V[]{X+Y} = \V[]{X} + \V[]{Y}$~\cite{Lipschutz2011Ch3}.
    \begin{equation}
        \label{eq:distributed variance for p}
        \V[]{\widehat{p}} = \frac{1}{N_U^2} \sum_{r=1}^{N_U} \V[]{\widehat{\mathcal{X}_{U_r}}}
    \end{equation}
    We apply Theorem \ref{thm: eves law} to $\V[]{\widehat{\mathcal{X}_{U_r}}}$,
    \begin{equation}
        \begin{split}
        \label{eq: variance decomposition of TrXU}
        \V[]{\widehat{\mathcal{X}_{U_r}}} &= \E[U_r]{\Vshot{\widehat{\mathcal{X}_{U_r}}}{U_r}} + \\
        & \V[U_r]{\Eshot{\widehat{\mathcal{X}_{U_r}}}{U_r}}
    \end{split}
    \end{equation}
    The second term can be further simplified since by Proposition~\ref{prp:unbiased estimator for TrXU}
    \[
        \Eshot{\widehat{\mathcal{X}_{U_r}}}{U_r} = \mathcal{X}_{U_r}
    \]
    Plugging this back into Eq.~\eqref{eq: variance decomposition of TrXU}, we get
    \[
    \V[]{\widehat{\mathcal{X}_{U_r}}} = \E[U_r]{\Vshot{\widehat{\mathcal{X}_{U_r}}}{U_r}} + \V[U_r]{\mathcal{X}_{U_r}}
    \]
    Since each of the $U_r$ is drawn i.i.d.\ from the same distribution, we have, for all $r \in \left\{ 1, \dots, N_U \right\}$,
    \[
    \E[U_r]{\Vshot{\widehat{\mathcal{X}_{U_r}}}{U_r}} = \E[U]{\Vshot{\widehat{\mathcal{X}_{U}}}{U}}
    \]
    \[
     \V[U_r]{\mathcal{X}_{U_r}} =  \V[U]{\mathcal{X}_{U}}
    \]
    There are $N_U$ terms in the sum in Eq.~\eqref{eq:distributed variance for p}, hence
    \[
    \begin{split}
        \V[]{\widehat{p}} = \frac{1}{N_U} \left(\V[U]{\mathcal{X}_{U}} + \E[U]{\Vshot{\widehat{\mathcal{X}_{U}}}{U}}\right)
    \end{split}
    \]
\end{proof}

\begin{proposition}
    The shot variance of $\widehat{\mathcal{X}_{U}}$ under a given unitary $U$ can be written as
    \[
    \Vshot{\widehat{\mathcal{X}_{U}}}{U} = \frac{4(N_M-2)\,\zeta_1 (U) + 2\,\zeta_2 (U)}{(N_M)^{\underline{2}}}
    \]
    where
    \[
    \begin{aligned}
    \zeta_1(U) &= 4^{N_A} \sum_{s_1, s_2, s_3} W_{s_1 s_2}\, W_{s_1 s_3}\, P_U(s_1) P_U(s_2) P_U(s_3) \\
    & - \mathcal{X}_U^2,\\
    \zeta_2(U) &= 4^{N_A} \sum_{s_1, s_2} W_{s_1 s_2}^2\, P_U(s_1) P_U(s_2) - \mathcal{X}_U^2.
    \end{aligned}
    \]
\end{proposition}

\begin{proof}
    First we have 
    \[
    \Vshot{\widehat{\mathcal{X}_{U}}}{U} = \frac{4(N_M-2)\,\zeta_1 (U) + 2\,\zeta_2 (U)}{(N_M)^{\underline{2}}}
    \]
    where $\zeta_1 (U) = \Covshot{2^{N_A} W_{s_1s_2}}{2^{N_A} W_{s_1s_3}}{U}$ and $\zeta_2 (U)= \Vshot{2^{N_A} W_{s_1s_2}}{U}$. This is just the degree-2 special case of Eq. (5.13) in Hoeffding's U-statistics paper~\cite{Hoeffding1948}. Hence, we just need to show that the $\zeta_1(U)$ and $\zeta_2 (U)$ can be written in the desired form in the statement of this proposition.

    Using the definition of covariance, we expand
    \[
    \begin{split}
        \zeta_1(U) 
        &= \Covshot{2^{N_A} W_{s_1s_2}}{2^{N_A} W_{s_1s_3}}{U} \\
        &= 4^{N_A} \Covshot{W_{s_1s_2}}{W_{s_1s_3}}{U} \\
        &= 4^{N_A} \bigl(\Eshot{W_{s_1s_2} W_{s_1s_3}}{U} - \\
        & \quad \Eshot{W_{s_1s_2}}{U} \Eshot{W_{s_1s_3}}{U}\bigr) \\
        &= 4^{N_A} \Biggl(\sum_{s_1, s_2, s_3} P_U(s_1) P_U(s_2) P_U(s_3) W_{s_1s_2} W_{s_1s_3} - \\
        & \quad \left(\sum_{s_1, s_2} P_U(s_1) P_U(s_2) W_{s_1s_2}\right) \\
        & \quad \left(\sum_{s_1, s_3} P_U(s_1) P_U(s_3) W_{s_1s_3}\right)\Biggr) \\
        &= 4^{N_A} \left(\sum_{s_1, s_2, s_3} W_{s_1 s_2}\, W_{s_1 s_3}\, P_U(s_1) P_U(s_2) P_U(s_3)\right) \\
        & \quad - \mathcal{X}_U^2
    \end{split}
    \]
    where in the last step we used Eq.~\eqref{eq: XU in terms of probability}.

    Next, we similarly expand the variance
    \[
    \begin{split}
        \zeta_2(U) 
        &= \Vshot{2^{N_A} W_{s_1s_2}}{U} \\
        &= 4^{N_A} \Vshot{W_{s_1s_2}}{U} \\
        &= 4^{N_A} \left(\Eshot{W_{s_1s_2}^2}{U} - \Eshot{W_{s_1s_2}}{U}^2\right) \\
        &= 4^{N_A} \left(\sum_{s_1, s_2} P_U(s_1) P_U(s_2) W_{s_1s_2}^2\right) - \mathcal{X}_U^2
    \end{split}
    \]
    Again, in the last step we used Eq.~\eqref{eq: XU in terms of probability}.
\end{proof}

\section{Unbiased Estimators of Quantities in Adaptive Decision}\label{app:Estimators of Quantities in Adaptive Decision}

\begin{lemma}\label{lem: summing over two indices of a product of Hamming kernels with one shared index}
    The sum over pairwise distinct shot indices of a product of two Hamming kernels that share one index can be written in count space as
    \[
    \begin{split}
    \sum_{\substack{j,l \,|\, i,j,l \\ \text{pairwise distinct}}} W_{s_i s_j} W_{s_i s_l} &= \left( \left( W \vb{c}_U \right)_{s_i} - 1 \right)^2 \\
    &\quad - \left( W_2 \vb{c}_U \right)_{s_i} + 1,
    \end{split}
    \]
    where $W_2$ is the entrywise square of the Hamming kernel $W$.
\end{lemma}

\begin{proof}
    We first consider a simpler case $\sum_{j \,|\, j \neq i} W_{s_i s_j}$. Since the sum runs over the measured string of each shot, we can rewrite it as a sum over the count of every string. Using
    \[
    \sum_{j} W_{s_i s_j} = \sum_{s} W_{s_i s} c_U(s),
    \]
    we can remove the self-pairing at the $i^{\text{th}}$ shot to get
    \begin{equation}
        \label{eq:summing a single index of Hamming kernel in count space}
        \sum_{j \,|\, j \neq i} W_{s_i s_j} = \sum_{s} W_{s_i s} c_U(s) - 1 = \left( W \vb{c}_U \right)_{s_i} - 1
    \end{equation}
    Similarly, 
    \[
    \sum_{j \,|\, j \neq i} W_{s_i s_j}^2 = \sum_{s} W_{s_i s}^2 c_U(s) - 1 = \left( W_2 \vb{c}_U \right)_{s_i} - 1
    \]
    Now consider $\left( \sum_{j \,|\, j \neq i} W_{s_i s_j} \right)^2$. One way to write it is to take the square of the count space form in Eq.~\eqref{eq:summing a single index of Hamming kernel in count space}. We can also write it by expanding the sum
    \[
    \begin{split}
        \left( \sum_{j \,|\, j \neq i} W_{s_i s_j} \right)^2 &= \left( \sum_{j \,|\, j \neq i} W_{s_i s_j} \right) \left( \sum_{l \,|\, l \neq i} W_{s_i s_l} \right) \\
        &= \sum_{j,l \,|\, j \neq i, l \neq i} W_{s_i s_j} W_{s_i s_l}
    \end{split}
    \]
    We can split this sum into two disjoint cases. The first case is when $i,j,l$ are pairwise distinct, while the second case is when $j \neq i$, $l \neq i$, and $j = l$ are simultaneously true. Then
    \[
    \begin{split}
        & \sum_{j,l \,|\, j \neq i, l \neq i} W_{s_i s_j} W_{s_i s_l} \\
        & = \sum_{\substack{j,l \,|\, i,j,l \\ \text{pairwise distinct}}} W_{s_i s_j} W_{s_i s_l} + \sum_{j\,|\, j \neq i} W_{s_i s_j}^2 \\
    \end{split}
    \]
    Then, we obtain the count space formula
    \[
    \begin{split}
        & \sum_{\substack{j,l \,|\, i,j,l \\ \text{pairwise distinct}}} W_{s_i s_j} W_{s_i s_l} \\
        &=  \sum_{j,l \,|\, j \neq i, l \neq i} W_{s_i s_j} W_{s_i s_l} - \sum_{j\,|\, j \neq i} W_{s_i s_j}^2 \\
        & = \left( \sum_{j \,|\, j \neq i} W_{s_i s_j} \right)^2 - \sum_{j\,|\, j \neq i} W_{s_i s_j}^2 \\
        & = \left( \left( W \vb{c}_U \right)_{s_i} - 1 \right)^2 - \left( W_2 \vb{c}_U \right)_{s_i} + 1
    \end{split}
    \]
\end{proof}

\begin{proposition}\label{prp:unbiased estimator for Chi_U2}
    The unbiased estimator for $\mathcal{X}_U^2$ can be written as in Eq.~\eqref{eq:estimator for TrXU^2 count space}, i.e.
    \[
    \widehat{\mathcal{X}_{U}^{\,2}} = \frac{4^{N_A}}{(N_M)^{\underline{4}}} \left[A_U^2 - 4T_U - 2\left(\vb{c}_U^{\top} W_2 \vb{c}_U - N_M\right)\right]
    \]
    in the sense that $\Eshot{\widehat{\mathcal{X}_{U}^{\,2}}}{U} = \mathcal{X}_U^{\,2}$, where
    \[
    A_U \equiv \vb{c}_U^{\top} W \vb{c}_U - N_M
    \]
    and 
    \[
    T_U \equiv \sum_s c_U(s)\,\left[ \left( \left( W \vb{c}_U \right)_{s} - 1 \right)^2 - \left( W_2 \vb{c}_U \right)_{s} + 1 \right]
    \]
\end{proposition}

\begin{proof}
    We first consider how to write the unbiased estimator of $\mathcal{X}_U^2$ in terms of the Hamming kernel. By Eq.~\eqref{eq: XU in terms of probability}, we know that $\mathcal{X}_U^2 = 4^{N_A} \left( \vb{p}_U^{\top} W \vb{p}_U \right)^2$. Then, we need a sum over Hamming kernels that will give us $\left( \vb{p}_U^{\top} W \vb{p}_U \right)^2$. Consider $\Eshot{W_{s_i s_j} W_{s_m s_n}}{U}$. If the two Hamming kernels are independent, we can write the expectation of their product as a product of their respective expectations, and by Lemma~\ref{lem:shot-expectation-Hamming-kernel} that should give us a $\left( \vb{p}_U^{\top} W \vb{p}_U \right)^2$. When the shot indices $i,j,m,n$ are pairwise distinct, the Hamming kernels $W_{s_i s_j}$ and $W_{s_m s_n}$ are indeed independent, so we have
    \[
    \begin{split}
        & \Eshot{\sum_{\substack{i,j,m,n \,|\, i,j,m,n \\ \text{pairwise distinct}}} W_{s_i s_j} W_{s_m s_n}}{U} \\
        & = \sum_{\substack{i,j,m,n \,|\, i,j,m,n \\ \text{pairwise distinct}}} \Eshot{W_{s_i s_j}}{U} \Eshot{W_{s_m s_n}}{U} \\
        & = \sum_{\substack{i,j,m,n \,|\, i,j,m,n \\ \text{pairwise distinct}}} \left( \vb{p}_U^{\top} W \vb{p}_U \right)^2 \\
        & = (N_M)^{\underline{4}} \left( \vb{p}_U^{\top} W \vb{p}_U \right)^2
    \end{split}
    \]
    where in the last step we used that $\left( \vb{p}_U^{\top} W \vb{p}_U \right)^2$ is independent of shot index, so each term in the sum is the same. The number of terms is the number of ways to choose an ordered 4-tuple from $N_M$ objects, which equals $(N_M)^{\underline{4}}$. Then,
    \[
    \begin{split}
        & \Eshot{\frac{4^{N_A}}{(N_M)^{\underline{4}}} \sum_{\substack{i,j,m,n \,|\, i,j,m,n \\ \text{pairwise distinct}}} W_{s_i s_j} W_{s_m s_n}}{U} \\
        & = 4^{N_A} \left( \vb{p}_U^{\top} W \vb{p}_U \right)^2 \\
        & = \mathcal{X}_U^2
    \end{split}
    \]
    Now we just need to write this Hamming kernel expression in terms of counts. First note that $A_U$ can be written in terms of the Hamming kernel as $A_U = \sum_{i,j\,|\, i \neq j} W_{s_i s_j}$. To show this, we sum over one of the indices first and apply Eq.~\eqref{eq:summing a single index of Hamming kernel in count space}
    \begin{equation}
        \label{eq:A_U in terms of Hamming kernel}
        \begin{split}
        \sum_{i,j\,|\, i \neq j} W_{s_i s_j} & = \sum_i \sum_{j\,|\, j \neq i} W_{s_i s_j} \\
        & = \sum_i \left[\left( W \vb{c}_U \right)_{s_i} - 1\right] \\
        & = \sum_{s_i} c_U(s_i) \left( \left( W \vb{c}_U \right)_{s_i} - 1 \right) \\
        & = \vb{c}_U^{\top} W \vb{c}_U - N_M \\
        & = A_U
    \end{split}
    \end{equation}
    To obtain the count-space expression of $\widehat{\mathcal{X}_{U}^{\,2}}$, we use an approach similar to that of Lemma~\ref{lem: summing over two indices of a product of Hamming kernels with one shared index}. Squaring a sum of Hamming kernels gives a sum over products of Hamming kernels, and expanding this sum produces the pairwise-distinct term $\sum_{\substack{i,j,m,n \,|\, i,j,m,n \\ \text{pairwise distinct}}} W_{s_i s_j} W_{s_m s_n}$. We take the square of $A_U$:
    \[
    A_U^2 = \left( \sum_{i,j\,|\, i \neq j} W_{s_i s_j} \right)^2  = \sum_{i,j,m,n\,|\, i \neq j, m \neq n} W_{s_i s_j} W_{s_m s_n}
    \]
    This sum can be split into three disjoint cases. In the first case, $i,j,m,n$ are pairwise distinct, which gives exactly the term we want. In the second case, exactly one index repeats; this includes the four subcases $i=m$, $i=n$, $j=m$, and $j=n$. In the third case, exactly two indices repeat; this includes the two subcases $\left(i=m \right)\land \left(j=n \right)$ and $\left(i=n \right)\land \left(j=m \right)$.

    For the second case, consider the subcase $i=m$ as an example; the relevant sum is $\sum_{\substack{i,j,n \,|\, i,j,n \\ \text{pairwise distinct}}} W_{s_i s_j} W_{s_i s_n}$. Summing two of the indices first recovers the form in Lemma~\ref{lem: summing over two indices of a product of Hamming kernels with one shared index}:
    \begin{equation}
        \label{eq: T_U count space Hamming kernel conversion}
    \begin{split}
        & \sum_{\substack{i,j,n \,|\, i,j,n \\ \text{pairwise distinct}}} W_{s_i s_j} W_{s_i s_n} \\
        & = \sum_{i} \sum_{\substack{j,n \,|\, i,j,n \\ \text{pairwise distinct}}} W_{s_i s_j} W_{s_i s_n} \\
        & = \sum_i \left[ \left( \left( W \vb{c}_U \right)_{s_i} - 1 \right)^2 - \left( W_2 \vb{c}_U \right)_{s_i} + 1 \right] \\
        & = \sum_{s_i} c_U(s_i)\,\left[ \left( \left( W \vb{c}_U \right)_{s_i} - 1 \right)^2 - \left( W_2 \vb{c}_U \right)_{s_i} + 1 \right] \\
        & = T_U
    \end{split}
    \end{equation}
    Using the symmetry of the Hamming kernel, the other three subcases can be brought into the same form as the summation above, so their contributions are all equal to $T_U$. In total, the case with exactly one repeated index gives $4T_U$.

    For the third case, consider the subcase $\left(i=m \right)\land \left(j=n \right)$ as an example; the relevant sum is $\sum_{i,j\,|\, i \neq j} W_{s_i s_j}^2$. By the same procedure as in Eq.~\eqref{eq:A_U in terms of Hamming kernel}, we obtain
    \[
    \sum_{i,j\,|\, i \neq j} W_{s_i s_j}^2 =  \vb{c}_U^{\top} W_2 \vb{c}_U - N_M
    \]
    By the same symmetry of the Hamming kernel, the other subcase contributes equally, so the total contribution is $2\left(\vb{c}_U^{\top} W_2 \vb{c}_U - N_M \right)$.

    Combining all three cases, we get
    \[
    \begin{split}
        A_U^2 &= \sum_{\substack{i,j,m,n \,|\, i,j,m,n \\ \text{pairwise distinct}}} W_{s_i s_j} W_{s_m s_n} \\
        &\quad + 4T_U + 2\left(\vb{c}_U^{\top} W_2 \vb{c}_U - N_M \right)
    \end{split}
    \]
    which implies
    \[
    \begin{split}
        & \frac{4^{N_A}}{(N_M)^{\underline{4}}} \sum_{\substack{i,j,m,n \,|\, i,j,m,n \\ \text{pairwise distinct}}} W_{s_i s_j} W_{s_m s_n} \\
        & = \frac{4^{N_A}}{(N_M)^{\underline{4}}} \left( A_U^2 - 4T_U - 2\left(\vb{c}_U^{\top} W_2 \vb{c}_U - N_M \right) \right)
    \end{split}
    \]
    giving exactly the desired count-space expression.
\end{proof}

\begin{theorem}
    An exactly unbiased estimator for unitary contribution to total variance $V_U$ can be written as in Eq.~\eqref{eq:V_U estimator}, i.e.
    \[
    \widehat{V_U}
    = \frac{1}{N_U}\sum_{r=1}^{N_U} \widehat{\mathcal{X}_{U_r}^{\,2}}
    - \frac{1}{(N_U)^{\underline{2}}} \sum_{r \neq r'} \widehat{\mathcal{X}_{U_r}}\;\widehat{\mathcal{X}_{U_{r'}}}.
    \]
    in the sense that
    \[
    \E[U_1, \dots, U_{N_U}]{\Eshot{\widehat{V_U}}{U_1 \dots U_{N_U}}} = V_U
    \]
\end{theorem}

\begin{proof}
    We first take the shot average
    \[
    \begin{split}
        & \Eshot{\widehat{V_U}}{U_1 \dots U_{N_U}} \\
        & = \frac{1}{N_U}\sum_{r=1}^{N_U} \Eshot{\widehat{\mathcal{X}_{U_r}^{\,2}}}{U_1 \dots U_{N_U}} - \\
        & \quad \frac{1}{(N_U)^{\underline{2}}} \sum_{r \neq r'} \Eshot{ \widehat{\mathcal{X}_{U_r}}\;\widehat{\mathcal{X}_{U_{r'}}}  }{U_1 \dots U_{N_U}} \\
        & = \frac{1}{N_U}\sum_{r=1}^{N_U} \Eshot{\widehat{\mathcal{X}_{U_r}^{\,2}}}{U_r} - \\
        & \quad \frac{1}{(N_U)^{\underline{2}}} \sum_{r \neq r'} \Eshot{ \widehat{\mathcal{X}_{U_r}} }{U_r} \Eshot{ \widehat{\mathcal{X}_{U_{r'}}} }{U_{r'}}\\
    \end{split}
    \]
    where in the last step we use that the measurement records obtained under different unitaries are independent, so that $\widehat{\mathcal{X}_{U_r}}$ and $\widehat{\mathcal{X}_{U_{r'}}}$ are independent for $r \neq r'$. We also use that $\Eshot{\widehat{\mathcal{X}_{U_r}^{\,2}}}{U_1 \dots U_{N_U}} =  \Eshot{\widehat{\mathcal{X}_{U_r}^{\,2}}}{U_r}$ and $\Eshot{\widehat{\mathcal{X}_{U_r}}}{U_1 \dots U_{N_U}} =  \Eshot{\widehat{\mathcal{X}_{U_r}}} {U_r}$ because keeping additional irrelevant unitaries constant does not affect the probability distribution. Then by Proposition~\ref{prp:unbiased estimator for TrXU} and Proposition~\ref{prp:unbiased estimator for Chi_U2}
    \[
    \begin{split}
        & \Eshot{\widehat{V_U}}{U_1 \dots U_{N_U}} \\
        & = \frac{1}{N_U}\sum_{r=1}^{N_U} \mathcal{X}_{U_r}^{\,2} - \frac{1}{(N_U)^{\underline{2}}} \sum_{r \neq r'} \mathcal{X}_{U_r} \mathcal{X}_{U_{r'}} 
    \end{split}
    \]
    Taking further average over the unitaries, we get 
    \[
    \begin{split}
        & \E[U_1, \dots, U_{N_U}]{\Eshot{\widehat{V_U}}{U_1 \dots U_{N_U}}}  \\
        & = \frac{1}{N_U}\sum_{r=1}^{N_U} \E[U_1, \dots, U_{N_U}]{\mathcal{X}_{U_r}^{\,2}} - \\ 
        & \quad \frac{1}{(N_U)^{\underline{2}}} \sum_{r \neq r'} \E[U_1, \dots, U_{N_U}]{\mathcal{X}_{U_r} \mathcal{X}_{U_{r'}}} \\
        & = \frac{1}{N_U}\sum_{r=1}^{N_U} \E[U_1, \dots, U_{N_U}]{\mathcal{X}_{U_r}^{\,2}} - \\ 
        & \quad \frac{1}{(N_U)^{\underline{2}}} \sum_{r \neq r'} \E[U_1, \dots, U_{N_U}]{\mathcal{X}_{U_r}} \E[U_1, \dots, U_{N_U}]{\mathcal{X}_{U_{r'}}} \\
        & = \frac{1}{N_U}\sum_{r=1}^{N_U} \E[U_r]{\mathcal{X}_{U_r}^{\,2}} - \frac{1}{(N_U)^{\underline{2}}} \sum_{r \neq r'} \E[U_{r}]{\mathcal{X}_{U_r}} \E[U_{r'}]{\mathcal{X}_{U_{r'}}} \\
        & = \E[U]{\mathcal{X}_{U}^{\,2}} - \E[U]{\mathcal{X}_{U}}^2
    \end{split}
    \]
    where in the last step we used that $\E[U_r]{\mathcal{X}_{U_r}^{\,2}} = \E[U]{\mathcal{X}_{U}^{\,2}}$ and $\E[U_r]{\mathcal{X}_{U_r}} = \E[U]{\mathcal{X}_{U}}$  for any $U_r$ since each $U_r$ is drawn independently and identically from the same distribution. Finally, by Eq.~\eqref{eq: V_U}, we have
    \[
    V_U \equiv \V[U]{\mathcal{X}_U} = \E[U]{\mathcal{X}_U^{\,2}} - \E[U]{\mathcal{X}_U}^2
    \]
    Hence,
    \[
    \E[U_1, \dots, U_{N_U}]{\Eshot{\widehat{V_U}}{U_1 \dots U_{N_U}}} = V_U
    \]
\end{proof}

\begin{theorem}
    An exactly unbiased estimator for $\zeta_1(U)$ defined in Eq.~\eqref{eq:zeta coefficients} can be written as
    \[
    \widehat{\zeta_1(U)} = \frac{4^{N_A}}{(N_M)^{\underline{3}}} T_U - \widehat{\mathcal{X}_{U}^{\,2}}
    \]
    in the sense that
    \[
    \Eshot{\widehat{\zeta_1(U)}}{U} = \zeta_1(U)
    \]
    where $T_U$ is defined in Eq.~\eqref{eq:T_U}
\end{theorem}

\begin{proof}
    We first show that the unbiased estimator can be written in Hamming kernel form as 
    \begin{equation}
        \label{eq: zeta1 estimator Hamming kernel form}
        \widehat{\zeta_1(U)} = \frac{4^{N_A}}{(N_M)^{\underline{3}}} \sum_{\substack{i,j,k \,|\, i,j,k \\ \text{pairwise distinct}}} W_{s_i s_j} W_{s_i s_k} - \widehat{\mathcal{X}_{U}^{\,2}}
    \end{equation}
    This can be done by directly taking the shot average, 
    \[
    \begin{split}
        & \Eshot{\widehat{\zeta_1(U)}}{U} \\
        & = \frac{4^{N_A}}{(N_M)^{\underline{3}}} \sum_{\substack{i,j,k \,|\, i,j,k \\ \text{pairwise distinct}}} \Eshot{W_{s_i s_j} W_{s_i s_k}}{U} \\
        & \quad - \Eshot{\widehat{\mathcal{X}_{U}^{\,2}}}{U} \\
        & = \frac{4^{N_A}}{(N_M)^{\underline{3}}} \sum_{\substack{i,j,k \,|\, i,j,k \\ \text{pairwise distinct}}} 
        \sum_{s_i, s_j, s_k} W_{s_i s_j} W_{s_i s_k} \\ 
        & \quad P_U(s_i) P_U(s_j) P_U(s_k) - \Eshot{\widehat{\mathcal{X}_{U}^{\,2}}}{U} \\
        & = 4^{N_A} \sum_{s_i, s_j, s_k} W_{s_i s_j} W_{s_i s_k} P_U(s_i) P_U(s_j) P_U(s_k) - \mathcal{X}_{U}^{\,2} \\
        & = \zeta_1(U)
    \end{split}
    \]
    where we used that given a fixed unitary, each shot is drawn independently and identically from the same ensemble, so $\Eshot{W_{s_i s_j} W_{s_i s_k}}{U}$ should be the same for every set of pairwise distinct indices $i,j,k$. Then, we just need to write the Hamming kernel expression in count space. By Eq.~\eqref{eq: T_U count space Hamming kernel conversion}, we get that 
    \[
    T_U = \sum_{\substack{i,j,n \,|\, i,j,n \\ \text{pairwise distinct}}} W_{s_i s_j} W_{s_i s_n}
    \]
    Substituting into Eq.~\eqref{eq: zeta1 estimator Hamming kernel form},
    \[
     \widehat{\zeta_1(U)} = \frac{4^{N_A}}{(N_M)^{\underline{3}}} T_U - \widehat{\mathcal{X}_{U}^{\,2}}
    \]
\end{proof}

\begin{theorem}
    An exactly unbiased estimator for $\zeta_2(U)$ defined in Eq.~\eqref{eq:zeta coefficients} can be written as
    \[
    \widehat{\zeta_2(U)} = \frac{4^{N_A}}{(N_M)^{\underline{2}}} \left(\vb{c}_{U}^{\top} W_2 \vb{c}_{U} - N_M\right) - \widehat{\mathcal{X}_{U}^{\,2}}
    \]
    in the sense that
    \[
    \Eshot{\widehat{\zeta_2(U)}}{U} = \zeta_2(U)
    \]
\end{theorem}

\begin{proof}
    Again, we first show that the Hamming kernel form
    \[
    \widehat{\zeta_2(U)} = \frac{4^{N_A}}{(N_M)^{\underline{2}}}\sum_{i,j \,|\, i \neq j} W_{s_is_j}^2 - \widehat{\mathcal{X}_{U}^{\,2}}
    \]
    is an unbiased estimator of $\zeta_2(U)$
    \[
    \begin{split}
        & \Eshot{\widehat{\zeta_2(U)}}{U} \\
        & = \frac{4^{N_A}}{(N_M)^{\underline{2}}} \sum_{i,j \,|\, i \neq j} \Eshot{W_{s_is_j}^2}{U} -  \Eshot{\widehat{\mathcal{X}_{U}^{\,2}}}{U} \\
        & = \frac{4^{N_A}}{(N_M)^{\underline{2}}} \sum_{i,j \,|\, i \neq j} \sum_{s_i, s_j} W_{s_is_j}^2 P_U(s_i) P_U(s_j) - \mathcal{X}_{U}^{\,2} \\
        & = 4^{N_A} \sum_{s_i, s_j} W_{s_is_j}^2 P_U(s_i) P_U(s_j) - \mathcal{X}_{U}^{\,2} \\
        & = \zeta_2(U)
    \end{split}
    \]
\end{proof}

\begin{corollary}
    Unbiased estimators for $\bar\zeta_1$ and $\bar\zeta_2$ can be written as
    \begin{equation}
        \widehat{\bar\zeta_i} = \frac{1}{N_U}\sum_{r=1}^{N_{U}} \widehat{\zeta_i(U_r)}
    \end{equation}
    in the sense that $\E[U_1, \dots, U_{N_U}]{\Eshot{\widehat{\bar\zeta_i}}{U_1 \dots U_{N_U}}} = \bar\zeta_i$ for $i=1,2$
\end{corollary}

\begin{proof}
    \[
    \begin{split}
        & \E[U_1, \dots, U_{N_U}]{\Eshot{\widehat{\bar\zeta_i}}{U_1 \dots U_{N_U}}} \\
        & = \frac{1}{N_U}\sum_{r=1}^{N_{U}}  \E[U_1, \dots, U_{N_U}]{\Eshot{\widehat{\zeta_i(U_r)}}{U_1 \dots U_{N_U}}} \\
        & = \frac{1}{N_U}\sum_{r=1}^{N_{U}}  \E[U_r]{\Eshot{\widehat{\zeta_i(U_r)}}{U_r}} \\
        & = \frac{1}{N_U} \sum_{r=1}^{N_{U}} \E[U_r]{ \zeta_i(U_r) } \\
        & = \frac{1}{N_U} \sum_{r=1}^{N_{U}} \bar\zeta_i \\
        & = \bar\zeta_i
    \end{split}
    \]
\end{proof}

\end{document}